\documentclass[aps,nofootinbib,twocolumn,showpacs]{revtex4}
\usepackage{amsmath,amsfonts,amssymb}
\usepackage{enumitem}
\usepackage{multi row,graphicx}
\usepackage[caption=false]{subfig}

\usepackage{array}
\newcolumntype{C}{>{\centering\arraybackslash}p{8ex}}

\begin{document}

\title{The Fast and the Fiducial:\\Augmented kludge waveforms for detecting extreme-mass-ratio inspirals}
\author{Alvin J. K. Chua}
\email{ajkc3@ast.cam.ac.uk}
\affiliation{Institute of Astronomy, University of Cambridge, Madingley Road, Cambridge CB3 0HA, United Kingdom}
\author{Christopher J. Moore}
\email{cjm96@cam.ac.uk}
\affiliation{Department of Applied Mathematics and Theoretical Physics, University of Cambridge, Wilberforce Road, Cambridge CB3 0WA, United Kingdom}
\author{Jonathan R. Gair}
\email{j.gair@ed.ac.uk}
\affiliation{School of Mathematics, University of Edinburgh, King's Buildings, Edinburgh EH9 3JZ, United Kingdom}
\date{\today}

\begin{abstract}
The extreme-mass-ratio inspirals (EMRIs) of stellar-mass compact objects into massive black holes are an important class of source for the future space-based gravitational-wave detector LISA. Detecting signals from EMRIs will require waveform models that are both accurate and computationally efficient. In this paper, we present the latest implementation of an augmented analytic kludge (AAK) model, publicly available at \href{https://github.com/alvincjk/EMRI_Kludge_Suite}{github.com/alvincjk/EMRI_Kludge_Suite} as part of an EMRI waveform software suite. This version of the AAK model has improved accuracy compared to its predecessors, with two-month waveform overlaps against a more accurate fiducial model exceeding 0.97 for a generic range of sources; it also generates waveforms 5--15 times faster than the fiducial model. The AAK model is well suited for scoping out data analysis issues in the upcoming round of mock LISA data challenges. A simple analytic argument shows that it might even be viable for detecting EMRIs with LISA through a semi-coherent template bank method, while the use of the original analytic kludge in the same approach will result in around 90\% fewer detections.
\end{abstract}

\pacs{04.30.-w, 04.70.Bw, 04.80.Nn, 95.55.Ym}
\maketitle

\section{Introduction}

The key sources for future space-based gravitational-wave (GW) detectors operating in the millihertz frequency band will include the inspirals of stellar-mass compact objects---typically stellar-origin black holes, but also potentially neutron stars or white dwarfs---into massive black holes (MBHs) at the centres of galaxies. Such systems are called extreme-mass-ratio inspirals (EMRIs), since the typical mass of the inspiralling object is $\sim10M_\odot$ while that of the central MBH is $\sim10^4$--$10^7M_\odot$ (to give emission at millihertz frequencies). EMRIs occur because an MBH is typically surrounded by clusters of stars, and various dynamical processes (including two-body scattering, tidal splitting of binaries and stripping of giant stars) can lead to compact objects being captured by and subsequently inspiralling into the MBH (see \cite{AEA2007,A2012} for comprehensive discussions of the astrophysical channels leading to EMRI formation).

In 2013, ESA identified GW detection from space \cite{AEA2013} as the science theme to be addressed by the third L-class mission (L3) in its Cosmic Vision scientific programme, with a provisional launch date of 2034. A call for mission proposals to address this science theme was issued in October 2016, and ESA has now selected a three-satellite interferometry mission---the Laser Interferometer Space Antenna (LISA) \cite{AEA2017}. EMRI detection formed a key part of both the L3 science case \cite{AEA2013} and the specification of the L3 mission requirements \cite{AEA2017}. Between a few and a few hundred EMRI observations are expected over the mission lifetime \cite{GEA2004,G2009,BCCG2016,BEA2017}, and these have tremendous potential for the purposes of astrophysics, cosmology and fundamental physics.

The observation of as few as ten EMRIs can provide a measurement of the slope of the black-hole mass function to better precision than is currently known \cite{GTV2010}. EMRIs can also be used as standard sirens to investigate the expansion history of the Universe, and hence to constrain cosmological parameters \cite{MH2008}. Finally, EMRI observations provide exquisite probes of gravitational physics; in particular, they can be used to map out the spacetime structure around the central MBH to high precision, allowing us to test if that geometry is described by general relativity or an alternative theory (see e.g. \cite{MCG2017}). We refer the reader to \cite{GVLB2013} for a comprehensive review of testing relativity with EMRIs and other LISA sources.

The scientific objectives described above will only be achieved if EMRIs can be successfully identified and characterised in the noisy LISA data stream. This is difficult because EMRI orbits are generally expected to be both eccentric and inclined to the equatorial plane of the MBH, such that the emitted GWs contain a complex superposition of three fundamental frequencies. Since an EMRI waveform depends on 14 different parameters and $\sim10^5$ waveform cycles can be observed in the LISA frequency band, there are a huge number of independent EMRI signals that must be searched for in the data.

Early work \cite{GEA2004} showed in theory that EMRI data analysis is possible using a semi-coherent approach, without actually demonstrating the effectiveness of such a method in practice. Between 2006 and 2011, the feasibility of LISA data analysis in general was explored through a sequence of mock LISA data challenges \cite{AEA2006,AEA2007a,BEA2008,BEA2008a,BEA2010}. These showed in practice that it is possible to correctly identify individual EMRIs in data sets without other sources, and using narrow parameter priors \cite{BEA2008,BEA2008a,BEA2010,BGP2009}. The successful approaches typically used techniques such as Markov chain Monte Carlo to stochastically explore the EMRI parameter space and find the best-fit parameter values.

Much work is still needed to move from these simple initial demonstrations to a practical suite of tools for EMRI characterisation. However, one thing that all of these data analysis techniques have in common is the need for models of EMRI signals that might be present in the data. These models need to be sufficiently faithful to astrophysical EMRI signals in order to identify them in the data, but also computationally inexpensive enough that they can be generated in the large numbers required for grid-based or stochastic searches.

EMRI waveforms can be modelled accurately using black-hole perturbation theory, which exploits the extreme mass ratio to describe the inspiral through expansions in mass ratio. Perturbation theory can be used to construct the gravitational self-force acting on the inspiralling object; these calculations are challenging both theoretically and computationally (see \cite{B2009,PPV2011} for recent reviews). The self-force at leading order in mass ratio is well understood and has been successfully computed, but tracking the phase of an EMRI accurately over $\sim10^5$ cycles requires second-order self-force calculations, which have yet to be performed (although the theory of such calculations has recently been worked out \cite{P2015}). Additional complicating features such as transient resonances \cite{BCCG2016,FH2012} also require more work to fully understand.

While fully consistent self-force waveforms for EMRIs will probably be available by the time that LISA is taking data, these are also likely to be extremely computationally intensive. Hence there is a clear need for computationally efficient, but faithful, models to use in data analysis. Two such ``kludge'' models for generic inspirals exist in the literature; these were constructed to scope out LISA data analysis issues, e.g. EMRI event rates, parameter estimation precision, etc. The ``analytic kludge'' (AK) model \cite{BC2004} uses the quadrupole emission from a Keplerian orbit \cite{PM1963} as its base, imposing on top of this relativistic effects such as precession of the periapsis and the orbital plane, and radiation-reaction-driven evolution of the orbital parameters. The AK model is very fast to evaluate, and was used as the reference model in past mock data challenges for this reason. However, it does not provide a good match to true EMRI signals (as gauged by comparison to perturbative calculations) and so is not appropriate for LISA data analysis.

The ``numerical kludge'' (NK) model builds the trajectory of an EMRI from an exact Kerr geodesic, the parameters of which are then evolved using expressions derived from post-Newtonian (PN) expansions and fits to perturbative calculations \cite{GG2006}. A waveform is computed from the resulting trajectory by identifying the Boyer--Lindquist coordinates of the Kerr spacetime with flat-space spherical polar coordinates \cite{BEA2007}. NK waveforms show high fidelity with more accurate Teukolsky-based waveforms \cite{H2001,DH2006} and can readily be extended to include additional physical effects, such as conservative corrections to the evolution \cite{HG2009} or a prescription for the force acting on the orbit \cite{GEA2011,WEA2012}. These waveforms therefore satisfy the requirement of faithfulness to true EMRI signals, but they are more computationally expensive to generate than AK waveforms.

In this paper, we describe an ``augmented analytic kludge'' (AAK) model that uses information from the NK model to improve the faithfulness of AK waveforms without significantly increasing their computational cost. This is achieved by mapping the parameters of the AK model to match the frequencies of NK waveforms. The AAK model was first introduced in \cite{CG2015}; here we further improve the model, give details on the released implementation at \href{https://github.com/alvincjk/EMRI_Kludge_Suite}{github.com/alvincjk/EMRI_Kludge_Suite}, and discuss its implications for EMRI detection with LISA. AAK waveforms are much closer to astrophysical signals than AK waveforms, and so can detect/localise candidate signals in LISA data with sufficient sensitivity/accuracy such that follow-up parameter estimation with perturbative waveforms is feasible. They are also about an order of magnitude faster to evaluate than NK waveforms. Hence the AAK model is well suited for use in future mock data challenges, and possibly even in the final LISA data analysis infrastructure.

A brief overview of the AK and NK models is given in Sec.~\ref{sec:kludges}, while the AAK model is presented in Sec.~\ref{sec:AAK}. In Sec.~\ref{subsec:map}, we first introduce the Kerr fundamental frequencies and the parameter-space map they induce in the AK model, before describing the technical details of the AAK implementation in Sec.~\ref{subsec:implementation}. The performance of the augmented model is then compared to that of the original AK model in Sec.~\ref{subsec:benchmarking}, with the more accurate but slower NK model used as the benchmark for both. In Sec.~\ref{sec:application}, the application of the AAK model to EMRI detection with LISA is considered; we provide an analytic estimate for the threshold signal amplitude required in a semi-coherent search, and assess the viability of the AK and AAK models for real LISA data analysis.

\section{Kludge waveform models}\label{sec:kludges}

A kludge in the context of EMRI modelling is any approximate model that uses a combination of formalisms to generate waveforms quickly and extensively for data analysis. Kludge waveforms capture many qualitative features of more accurate EMRI waveforms, and (owing to their modular construction) can be modified to incorporate self-force information as it becomes available.

Two widely used kludges are introduced briefly in this section: the AK waveform of Barack \& Cutler \cite{BC2004}, which is very fast to compute and provides the basis for our new model, and the NK waveform of Babak et al. \cite{BEA2007}, which we take as a fiducial model for calibration and benchmarking purposes. Other approximate EMRI models exist but at varying levels of implementation (e.g. \cite{ST2003,FEH2016}), and we do not consider them in this work (apart from the PN fluxes of Sago \& Fujita \cite{SF2015}, which are used as part of the AAK model).

Assuming the spin of the compact object is negligible, an EMRI can be described by 14 parameters: the two masses $(\mu,M\gg\mu)$ of the system, the three components of the central black hole's spin vector $\mathbf{S}$, three constants $\mathbf{E}$ describing the compact object's (instantaneous) orbit, the three components of the compact object's position vector $\mathbf{X}$ with respect to the black hole, and the three components of the system's position vector $\mathbf{R}$ with respect to the Solar System.

Of these 14 degrees of freedom, seven are extrinsic to the source: two in $\mathbf{S}$ and one in $\mathbf{X}$ (corresponding to spatial rotation of the source), three in $\mathbf{R}$ (corresponding to spatial translation), and one in $\mathbf{E}$ (corresponding to temporal translation). The parameters of an EMRI model are often chosen to decouple the intrinsic degrees of freedom from the extrinsic ones, which are generally cheaper to search over during data analysis \cite{BCV2003}.

Schematically, the main ingredients of a kludge waveform model are then (i) the evolution of the orbital constants along the inspiral (i.e. the ``phase-space'' trajectory), using PN or fitted fluxes $\mathbf{F}$:
\begin{equation}
\dot{\mathbf{E}}=\mathbf{F}(\mu,M,\mathbf{S},\mathbf{E});
\end{equation}
(ii) the construction of the compact object's worldline (i.e. the ``configuration-space'' trajectory), using geodesic or flux-derived expressions $\mathbf{G}$:
\begin{equation}
\dot{\mathbf{X}}=\mathbf{G}(\mu,M,\mathbf{S},\mathbf{E});
\end{equation}
and (iii) the generation of the waveform field $h$ at the detector, using some weak-field multipole formula $H$:
\begin{equation}
h(t)=H(\mathbf{X},\mathbf{R}).
\end{equation}

\subsection{Analytic kludge}\label{subsec:AK}

In the AK model \cite{BC2004}, both the orbital trajectory and the waveform are computed in a flat-space approximation, with relativistic effects such as inspiralling and precession added separately. The trajectory is built out of rotating Keplerian ellipses. Radiation reaction is introduced in phase space, where the orbital constants describing a Keplerian ellipse are evolved with PN equations. In configuration space, the orientation of this ellipse is also evolved with PN equations to simulate relativistic precession. The waveform is then generated using the Peters--Mathews mode-sum approximation for Keplerian orbits \cite{PM1963}, in which the mass quadrupole moment is decomposed into harmonics of the Keplerian orbital frequency.

Since a Keplerian orbit is confined within the plane normal to its angular momentum vector $\mathbf{L}$, the AK waveform is constructed in an $\mathbf{L}$-based coordinate frame
\begin{equation}\label{eq:L_tilde-based}
(\hat{\mathbf{x}},\hat{\mathbf{y}},\hat{\mathbf{z}})_\mathbf{\tilde{L}}:=\left(\frac{(\hat{\mathbf{R}}\cdot\hat{\mathbf{L}})\hat{\mathbf{L}}-\hat{\mathbf{R}}}{\mathcal{S}_{\mathbf{L},\mathbf{R}}},\frac{\hat{\mathbf{R}}\times\hat{\mathbf{L}}}{\mathcal{S}_{\mathbf{L},\mathbf{R}}},\hat{\mathbf{L}}\right),
\end{equation}
and projected transverse to the wave frame
\begin{equation}\label{eq:AK_frame}
(\hat{\mathbf{x}},\hat{\mathbf{y}},\hat{\mathbf{z}})_\mathrm{AK}:=\left(\frac{\hat{\mathbf{R}}\times\hat{\mathbf{L}}}{\mathcal{S}_{\mathbf{L},\mathbf{R}}},\frac{\hat{\mathbf{L}}-(\hat{\mathbf{L}}\cdot\hat{\mathbf{R}})\hat{\mathbf{R}}}{\mathcal{S}_{\mathbf{L},\mathbf{R}}},-\hat{\mathbf{R}}\right),
\end{equation}
where the normalisation factor $\mathcal{S}_{\mathbf{L},\mathbf{R}}:=(1-(\hat{\mathbf{L}}\cdot\hat{\mathbf{R}})^2)^{1/2}$. These two frames are made time-varying (with respect to a fixed heliocentric and ecliptic-based frame \cite{C1998}) through the forced precession of $\mathbf{L}$.

The two waveform polarisations in the transverse--traceless gauge (with the usual $(+,\times)$ convention for $(\hat{\mathbf{x}},\hat{\mathbf{y}})_\mathrm{AK}$) are given by the $n$-mode sums
\begin{equation}\label{eq:mode_sum}
h_+=\sum_{n=1}^\infty{h^+_n},\quad h_\times=\sum_{n=1}^\infty{h^\times_n}
\end{equation}
with
\begin{equation}
h^+_n=(1+(\hat{\mathbf{R}}\cdot\hat{\mathbf{L}})^2)(b_n\sin{2\tilde{\gamma}}-a_n\cos{2\tilde{\gamma}})+(1-(\hat{\mathbf{R}}\cdot\hat{\mathbf{L}})^2)c_n,
\end{equation}
\begin{equation}
h^\times_n=2(\hat{\mathbf{R}}\cdot\hat{\mathbf{L}})(b_n\cos{2\tilde{\gamma}}+a_n\sin{2\tilde{\gamma}}),
\end{equation}
where $\tilde{\gamma}$ is an azimuthal angle in the orbital plane measuring the direction of periapsis with respect to $(\hat{\mathbf{R}}\cdot\hat{\mathbf{L}})\hat{\mathbf{L}}-\hat{\mathbf{R}}$ (i.e. the orthogonal projection of $\hat{\mathbf{z}}_\mathrm{AK}$ onto the plane normal to $\hat{\mathbf{L}}$).\footnote{We have changed some of the notation in \cite{BC2004} for consistency, since we are constructing a hybrid model using different formalisms. For example: the notation for the angles $\gamma$ and $\tilde{\gamma}$ has been swapped; the notation $\nu$ for the orbital frequency $\dot{\Phi}/(2\pi)$ is unused; the notation for the inclination $\lambda$ is now $\iota$.} The functions $(a_n,b_n,c_n)$ describe the changing mass quadrupole moment of a Keplerian orbit with mean anomaly $\Phi(t)$, eccentricity $e$ and orbital angular frequency $\dot{\Phi}$, and are given by \cite{PM1963}
\begin{eqnarray}\label{eq:PM_a}
a_n&=&-n\mathcal{A}(J_{n-2}(ne)-2eJ_{n-1}(ne)+(2/n)J_n(ne)\nonumber\\
&&+2eJ_{n+1}(ne)-J_{n+2}(ne))\cos{n\Phi},
\end{eqnarray}
\begin{eqnarray}\label{eq:PM_b}
b_n&=&-n\mathcal{A}(1-e^2)^{1/2}(J_{n-2}(ne)-2J_n(ne)\nonumber\\
&&+J_{n+2}(ne))\sin{n\Phi},
\end{eqnarray}
\begin{equation}\label{eq:PM_c}
c_n=2\mathcal{A}J_n(ne)\cos{n\Phi},
\end{equation}
where the $J_n$ are Bessel functions of the first kind, and $\mathcal{A}=(\dot{\Phi}M)^{2/3}\mu/|\mathbf{R}|$ in the extreme-mass-ratio limit.

In the ecliptic-based coordinate system, the sky position $\hat{\mathbf{R}}\equiv(\theta_S,\phi_S)$ of the source and the black-hole spin orientation $\hat{\mathbf{S}}\equiv(\theta_K,\phi_K)$ are effectively constant. It is convenient to represent $\hat{\mathbf{L}}$ in ecliptic coordinates with respect to $\hat{\mathbf{S}}$. We have
\begin{equation}\label{eq:L_ecliptic}
\hat{\mathbf{L}}=\hat{\mathbf{S}}\cos{\iota}+\left(\frac{\hat{\mathbf{z}}-(\hat{\mathbf{z}}\cdot\hat{\mathbf{S}})\hat{\mathbf{S}}}{|\hat{\mathbf{z}}-(\hat{\mathbf{z}}\cdot\hat{\mathbf{S}})\hat{\mathbf{S}}|}\cos{\alpha}+\frac{\hat{\mathbf{S}}\times\hat{\mathbf{z}}}{|\hat{\mathbf{S}}\times\hat{\mathbf{z}}|}\sin{\alpha}\right)\sin{\iota},
\end{equation}
where $\hat{\mathbf{z}}=[0,0,1]^T$ is normal to the ecliptic plane, $\iota$ is the inclination angle between $\hat{\mathbf{L}}$ and $\hat{\mathbf{S}}$, and $\alpha$ is an azimuthal angle in the spin-equatorial plane measuring the direction of $\hat{\mathbf{L}}-(\hat{\mathbf{L}}\cdot\hat{\mathbf{S}})\hat{\mathbf{S}}$ with respect to $\hat{\mathbf{z}}-(\hat{\mathbf{z}}\cdot\hat{\mathbf{S}})\hat{\mathbf{S}}$ (i.e. the angle between the orthogonal projections of $\hat{\mathbf{L}}$ and $\hat{\mathbf{z}}$ onto the plane normal to $\hat{\mathbf{S}}$). Furthermore, since $\tilde{\gamma}$ is neither intrinsic nor extrinsic, it is useful to define the purely intrinsic parameter $\gamma:=\tilde{\gamma}-\beta$, where $\beta=\beta(\hat{\mathbf{R}},\hat{\mathbf{S}},\hat{\mathbf{L}})=\beta(\theta_S,\phi_S,\theta_K,\phi_K,\iota,\alpha)$ is an azimuthal angle in the orbital plane measuring the direction of $\hat{\mathbf{L}}\times\hat{\mathbf{S}}$ with respect to $(\hat{\mathbf{R}}\cdot\hat{\mathbf{L}})\hat{\mathbf{L}}-\hat{\mathbf{R}}$.

Only one of the six parameters comprising $\mathbf{E}=(e,\iota,\dot{\Phi})$ and $\mathbf{X}=(\Phi(t),\gamma,\alpha)$ in the above Keplerian setup changes with time. In the AK model, $(e,\dot{\Phi},\gamma,\alpha)$ are promoted to functions of time and evolved with mixed-order PN expressions that depend on $(\mu,M,a=|\mathbf{S}|/M,\mathbf{E})$ \cite{BC1975,B1991,JS1992,R1996}, while $\iota$ is approximated as constant (since the inclination angle of a typical EMRI varies extremely slowly \cite{H2000}). The Keplerian orbit shrinks and circularises as $\dot{\Phi}(t)$ and $e(t)$ increase and decrease respectively. From \eqref{eq:L_ecliptic}, the time dependence of the orbital orientation $\hat{\mathbf{L}}(t)$ is confined to $\alpha(t)$, where $\dot{\alpha}$ is precisely the angular rate of Lense--Thirring precession. Finally, the angular rate of periapsis precession is given by $\dot{\gamma}+\dot{\alpha}$ since $\gamma(t)$ is measured with respect to $\hat{\mathbf{L}}(t)\times\hat{\mathbf{S}}$.

While the waveform field is effectively planar at the Solar System and may be calculated in the fixed heliocentric frame (as opposed to a detector-centric one), the rotational and orbital motion of LISA in the ecliptic plane must be factored into the detector's response to the GW. In the standard LISA framework, the waveform polarisations $h_{+,\times}$ are transformed into the response functions $h_{I,II}$ via
\begin{eqnarray}\label{eq:response_function}
h_I=\frac{\sqrt{3}}{2}(F_I^+h_++F_I^\times h_\times),\nonumber\\
h_{II}=\frac{\sqrt{3}}{2}(F_{II}^+h_++F_{II}^\times h_\times),
\end{eqnarray}
where the antenna pattern functions \cite{ACST1994}
\begin{eqnarray}\label{eq:antenna_Ip}
F_I^+&=&\frac{1}{2}(1+\cos^2{\theta_D})(\cos{2\phi_D})(\cos{2\psi_D})\nonumber\\
&&-(\cos{\theta_D})(\sin{2\phi_D})(\sin{2\psi_D}),
\end{eqnarray}
\begin{eqnarray}\label{eq:antenna_Ic}
F_I^\times&=&\frac{1}{2}(1+\cos^2{\theta_D})(\cos{2\phi_D})(\sin{2\psi_D})\nonumber\\
&&+(\cos{\theta_D})(\sin{2\phi_D})(\cos{2\psi_D}),
\end{eqnarray}
\begin{eqnarray}\label{eq:antenna_IIp}
F_{II}^+&=&\frac{1}{2}(1+\cos^2{\theta_D})(\sin{2\phi_D})(\cos{2\psi_D})\nonumber\\
&&+(\cos{\theta_D})(\cos{2\phi_D})(\sin{2\psi_D}),
\end{eqnarray}
\begin{eqnarray}\label{eq:antenna_IIc}
F_{II}^\times&=&\frac{1}{2}(1+\cos^2{\theta_D})(\sin{2\phi_D})(\sin{2\psi_D})\nonumber\\
&&-(\cos{\theta_D})(\cos{2\phi_D})(\cos{2\psi_D})
\end{eqnarray}
depend on the sky location $(\theta_D,\phi_D)$ and polarisation angle $\psi_D$ of the source in a detector-based coordinate system, and hence rotate with respect to $\hat{\mathbf{R}}$ as the plane of the detector along its orbit precesses around the ecliptic plane. Doppler modulation of the waveform phase (through $\Phi(t)$) is also included to correct for the orbital motion of the detector itself.

The AK model was the first waveform model used to investigate the precision of LISA parameter estimation over the full (modulo compact-object spin) EMRI parameter space \cite{BC2004}. Due to its computational efficiency, the model has also been employed in past mock LISA data challenges to generate injected signals in simulated data and parametrised templates for search algorithms \cite{BEA2008,BEA2008a,BEA2010,BGP2009}. However, the approximate waveforms it produces are demonstrably inaccurate, and will result in reduced detection and parameter estimation performance if used to analyse data sets containing realistic EMRI signals.

\subsection{Numerical kludge}\label{subsec:NK}

In the NK model \cite{BEA2007}, the orbital trajectory is computed in curved space with a treatment that is fully relativistic up to the evolution of orbital constants \cite{GHK2002,GG2006}, i.e. it is built out of Kerr geodesics. The three constants of motion for a geodesic are evolved with Teukolsky-fitted PN equations, which introduces radiation reaction. In configuration space, precession effects are obtained for free by integrating the geodesic equations along the phase-space trajectory. The curved-space coordinates of the compact object's worldline are then associated artificially with coordinates in flat space, and the waveform is generated using the standard quadrupole formula (or variants that include additional contributions from higher-order moments of mass \cite{B1973,P1977}).

The NK waveform is constructed in an $\mathbf{S}$-based coordinate frame
\begin{equation}\label{eq:S-based}
(\hat{\mathbf{x}},\hat{\mathbf{y}},\hat{\mathbf{z}})_\mathbf{S}:=\left(\frac{\hat{\mathbf{R}}\times\hat{\mathbf{S}}}{\mathcal{S}_{\mathbf{S},\mathbf{R}}},\frac{\hat{\mathbf{R}}-(\hat{\mathbf{R}}\cdot\hat{\mathbf{S}})\hat{\mathbf{S}}}{\mathcal{S}_{\mathbf{S},\mathbf{R}}},\hat{\mathbf{S}}\right),
\end{equation}
and projected transverse to the wave frame
\begin{equation}\label{eq:NK_frame}
(\hat{\mathbf{x}},\hat{\mathbf{y}},\hat{\mathbf{z}})_\mathrm{NK}:=\left(\frac{\hat{\mathbf{R}}\times\hat{\mathbf{S}}}{\mathcal{S}_{\mathbf{S},\mathbf{R}}},\frac{\hat{\mathbf{S}}-(\hat{\mathbf{S}}\cdot\hat{\mathbf{R}})\hat{\mathbf{R}}}{\mathcal{S}_{\mathbf{S},\mathbf{R}}},-\hat{\mathbf{R}}\right),
\end{equation}
where the normalisation factor $\mathcal{S}_{\mathbf{S},\mathbf{R}}:=(1-(\hat{\mathbf{S}}\cdot\hat{\mathbf{R}})^2)^{1/2}$. Aligning the $z$-axis with $\mathbf{S}$ is a more natural choice for the NK model, since the compact object's worldline is computed in Boyer--Lindquist coordinates. The two wave frames \eqref{eq:AK_frame} and \eqref{eq:NK_frame} are related by a (time-varying) rotation about $\mathbf{R}$.

Using the standard quadrupole formalism, the two waveform polarisations in the transverse--traceless gauge (with the $(+,\times)$ convention for $(\hat{\mathbf{x}},\hat{\mathbf{y}})_\mathrm{NK}$) are given by
\begin{equation}\label{eq:modes}
h_+=\frac{1}{2}h_{ij}H^+_{ij},\quad h_\times=\frac{1}{2}h_{ij}H^\times_{ij}
\end{equation}
with
\begin{equation}\label{eq:quadrupole}
h_{ij}=\frac{2}{|\mathbf{R}|}\left(P_{ik}P_{jl}-\frac{1}{2}P_{ij}P_{kl}\right)\ddot{I}_{kl},
\end{equation}
where $\ddot{I}_{ij}(t)$ is the second time derivative of the source's mass quadrupole moment $I_{ij}(t)$. The polarisation tensors $H^{+,\times}_{ij}$ and transverse projection tensor $P_{ij}$ are given by
\begin{equation}
H^+_{ij}=(\hat{\mathbf{x}}_i\hat{\mathbf{x}}_j-\hat{\mathbf{y}}_i\hat{\mathbf{y}}_j)_\mathrm{NK},
\end{equation}
\begin{equation}
H^\times_{ij}=(\hat{\mathbf{x}}_i\hat{\mathbf{y}}_j+\hat{\mathbf{y}}_i\hat{\mathbf{x}}_j)_\mathrm{NK},
\end{equation}
\begin{equation}
P_{ij}=(\delta_{ij}-\hat{\mathbf{z}}_i\hat{\mathbf{z}}_j)_\mathrm{NK},
\end{equation}
where $\delta_{ij}$ is the Kronecker delta.

In the extreme-mass-ratio limit, the mass quadrupole moment is simply
\begin{equation}\label{eq:quadrupole_moment}
I_{ij}=\mu x_ix_j,
\end{equation}
where the $x_i(t)$ are Cartesian components of the compact object's position vector $\mathbf{X}$ with respect to the frame \eqref{eq:S-based} centred on the black hole. Although \eqref{eq:quadrupole} (with \eqref{eq:quadrupole_moment}) is a weak-field equation in flat-space coordinates, the NK model specifies and calculates $(x_1,x_2,x_3)=(r\sin{\theta}\cos{\phi},r\sin{\theta}\sin{\phi},r\cos{\theta})$ in Boyer--Lindquist coordinates. The self-consistency of this approach clearly degrades further into the strong field, but does not severely impact the effectiveness of the NK waveforms as an approximation to Teukolsky-based ones \cite{BEA2007}.

A timelike Kerr geodesic is described fully by three first integrals of motion: the orbital energy $E$, the projection $L_z$ of the orbital angular momentum $\mathbf{L}$ onto $\mathbf{S}$, and the quadratic Carter constant $Q$. Along such an orbit, $(r(t),\theta(t),\phi(t))$ are obtained by integrating the geodesic equations for a test particle in the Kerr spacetime; these are written in canonical form as \cite{MTW1973}
\begin{equation}\label{eq:r_geodesic}
\Sigma\frac{dr}{d\tau}=\pm\sqrt{V_r},
\end{equation}
\begin{equation}\label{eq:theta_geodesic}
\Sigma\frac{d\theta}{d\tau}=\pm\sqrt{V_\theta},
\end{equation}
\begin{equation}\label{eq:phi_geodesic}
\Sigma\frac{d\phi}{d\tau}=V_\phi,
\end{equation}
\begin{equation}\label{eq:t_geodesic}
\Sigma\frac{dt}{d\tau}=V_t,
\end{equation}
where $\tau$ is proper time along the worldline and $\Sigma=r^2+a^2\cos^2\theta$. The potential functions $V_{r,\theta,\phi,t}$ are given by
\begin{equation}\label{eq:V_r}
V_r(r)=P^2-(r^2+(L_z-aE)^2+Q)\Delta,
\end{equation}
\begin{equation}\label{eq:V_theta}
V_\theta(\theta)=Q-\cos^2{\theta}\left(a^2(1-E^2)+\frac{L_z^2}{\sin^2{\theta}}\right),
\end{equation}
\begin{equation}\label{eq:V_phi}
V_\phi(r,\theta)=\frac{L_z}{\sin^2{\theta}}-aE+\frac{aP}{\Delta},
\end{equation}
\begin{equation}\label{eq:V_t}
V_t(r,\theta)=aL_z-a^2E\sin^2{\theta}+\frac{(r^2+a^2)P}{\Delta},
\end{equation}
with $P=E(r^2+a^2)-aL_z$ and $\Delta=r^2-2Mr+a^2$.

In practice, it is convenient to work with alternative parametrisations of $(E,L_z,Q)$. For a bound orbit, the geodesic may be specified by the parameters $(r_p,r_a,\theta_\mathrm{min})$ (the values of $r$ at periapsis and apoapsis, and the minimal value of $\theta$ respectively), which fully describe the range of motion in the radial and polar coordinates. The roots of $V_r$ determine $r_p$ and $r_a$, while the roots of $V_\theta$ determine $\cos{\theta_\mathrm{min}}$ (the maximal value of $\cos{\theta}$). Another parametrisation is $(e,\iota,p)$ (the quasi-Keplerian eccentricity, inclination and semi-latus rectum); these are defined in terms of $(r_p,r_a,\theta_\mathrm{min})$ as
\begin{equation}
(e,\iota,p):=\left(\frac{r_a-r_p}{r_a+r_p},\frac{\pi}{2}-\theta_\mathrm{min},\frac{2r_ar_p}{r_a+r_p}\right).
\end{equation}
Finally, since the configuration-space parameters $(r,\theta)$ oscillate between the bounds $r_p\leq r\leq r_a$ and $\theta_\mathrm{min}\leq\theta\leq\pi-\theta_\mathrm{min}$, it is useful to define
\begin{equation}\label{eq:NK_phases}
(\psi,\chi):=\left(\cos^{-1}{\left(\frac{p-r}{er}\right)},\cos^{-1}{\left(\frac{\cos{\theta}}{\cos{\theta_\mathrm{min}}}\right)}\right),
\end{equation}
such that $\psi$ (the quasi-Keplerian true anomaly) and $\chi$ are the phases of radial and polar motion respectively.

The orbital constants $\mathbf{E}=(E,L_z,Q)$ in the above geodesic setup do not vary with time. Radiation reaction is added to the NK model by evolving $\mathbf{E}$ with fluxes that depend on $(\mu,M,a,\mathbf{E})$ (note that the inclination $\iota(t)=\tan^{-1}(\sqrt{Q}/L_z)$ is correctly time-dependent in this model). These fluxes are mixed-order PN expressions that have been fitted to the results of Teukolsky-based computations for circular inclined orbits \cite{GG2006}. Integrating the geodesic equations along the phase-space trajectory then gives $\mathbf{X}=(\psi(t),\chi(t),\phi(t))$, complete with relativistic precession. Once the waveform polarisations $h_{+,\times}$ have been calculated via \eqref{eq:modes}--\eqref{eq:quadrupole_moment}, the LISA response functions $h_{I,II}$ may be obtained through the method outlined in Sec.~\ref{subsec:AK}.

Waveforms from the NK model display excellent agreement with Teukolsky-based waveforms in the strong-field regime; they are reliable up to a closest approach of $r_p\approx5M$, with typical matches of over 0.95 \cite{BEA2007}. NK waveforms might even be accurate enough to serve as templates in actual LISA detection algorithms. However, they are still slightly expensive to generate in large numbers due to the relatively elaborate construction of the phase- and configuration-space trajectories, while added computational cost also arises in the parameter conversion $(E,L_z,Q)\leftrightarrow(e,\iota,p)$, the handling of plunge, etc.

\section{Augmented analytic kludge}\label{sec:AAK}

\begin{figure}
\centering
\includegraphics[width=\columnwidth]{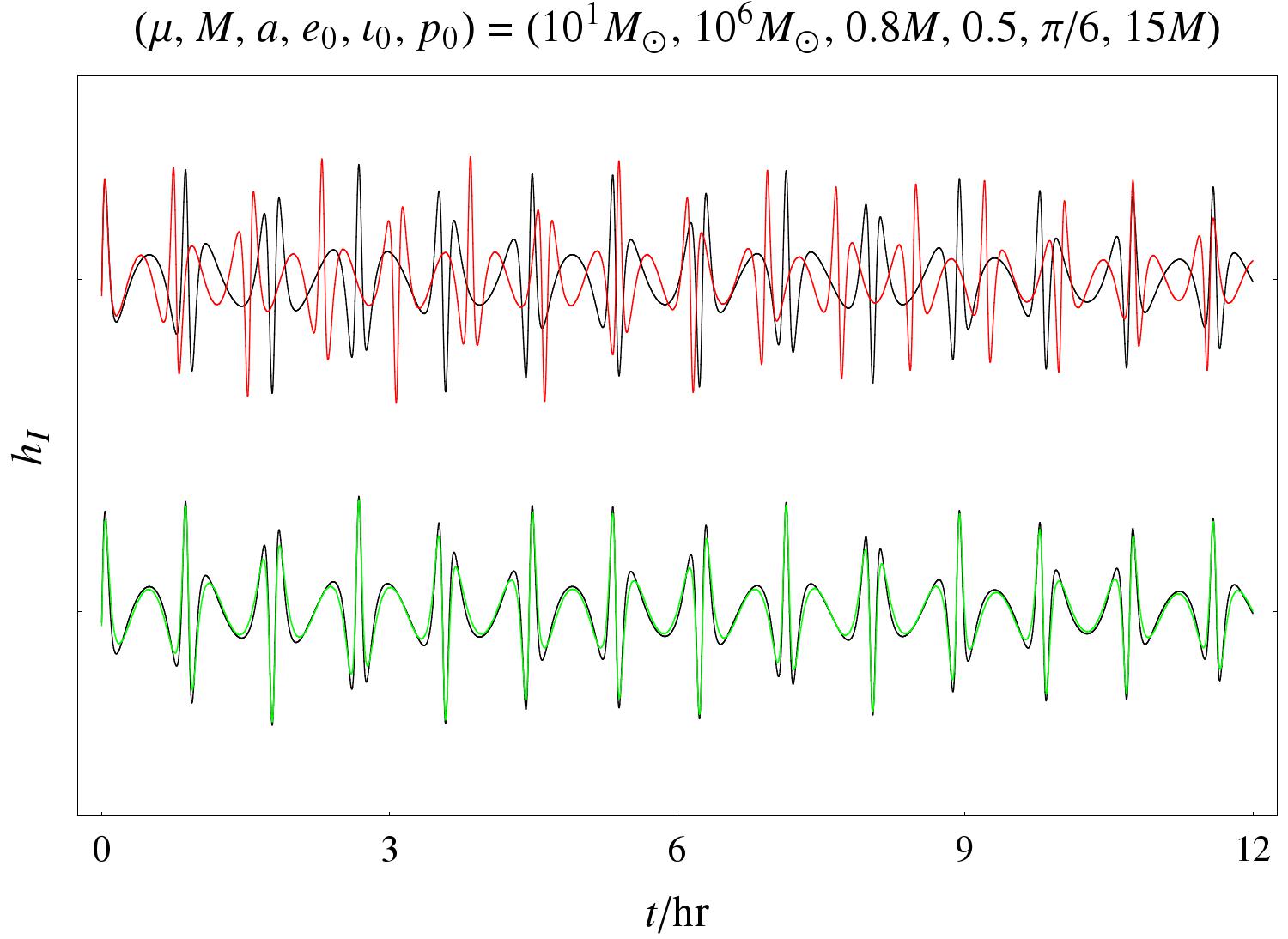}
\caption{First 12 hours of AK (red) and AAK (green) waveforms overlaid on NK waveform (black), for the early inspiral of a $(10^1,10^6)M_\odot$ EMRI with initial semi-latus rectum $p_0=15M$. Figure reproduced from \cite{CG2015}.}
\label{fig:waveform_comparison}
\end{figure}

The AK model is 5--15 times faster than the NK model at generating year-long waveforms sampled at 0.2\,Hz for a generic $(10^1,10^6)M_\odot$ EMRI with low initial eccentricity ($e_0\lesssim0.3$); this speed-up is enhanced for longer waveform durations, but diminished for higher initial eccentricity (since more modes must be summed in the Peters--Mathews approximation).\footnote{\label{foot:modes}The sums in \eqref{eq:mode_sum} must be truncated at some arbitrary number of modes $N$, which directly affects both the speed and accuracy of the AK model. This number may be specified by setting a threshold for the relative power radiated into the $N$-th harmonic, and has been experimentally determined to scale linearly with eccentricity \cite{BC2004}. We use $N=\lfloor30e_0\rfloor$ as the default value for both the AK and AAK models.} However, AK waveforms suffer from severe dephasing with respect to NK waveforms, even at the early-inspiral stage. In Fig.~\ref{fig:waveform_comparison}, the AK waveform for a $(10^1,10^6)M_\odot$ EMRI with initial semi-latus rectum $p_0=15M$ matches the qualitative features of the corresponding NK waveform, but is a full cycle out of phase within three hours. This is due to the mismatched frequencies in the two models.

In Secs~\ref{subsec:map} and \ref{subsec:implementation}, we describe the construction of a hybrid model that capitalises on the benefits of both kludges. The AK model is augmented with an initial map to the fundamental frequencies of Kerr geodesic motion, which corrects the instantaneous phasing as shown in Fig.~\ref{fig:waveform_comparison}. Over longer timescales, the mapped orbital trajectory is further improved through self-consistent PN evolution and a local polynomial fit to the phase-space trajectory of the NK model. Fast algorithms for higher-order fits and plunge handling have been incorporated in the latest implementation of the AAK model, which has been made publicly available at \href{https://github.com/alvincjk/EMRI_Kludge_Suite}{github.com/alvincjk/EMRI_Kludge_Suite} as part of a software suite for generating kludges.

The initial version of the AAK model yields waveforms that can remain phase-coherent with NK waveforms for over two months, but with overlap values lower than 0.97 \cite{CG2015}. This is the commonly chosen minimal match for a waveform template bank that corresponds to a $90\%$-ideal observed event rate \cite{O1996}, and thus ensures the equivalent localisation of any signal detected with such banks of AAK and NK templates. In Sec.~\ref{subsec:benchmarking}, we report further improved results for the present AAK implementation. Two-month overlaps higher than 0.97 are found for EMRIs with varying spin and eccentricity; however, the overlaps still degrade with proximity to plunge, due to the divergence of the AAK and NK trajectories deep within the strong field.

\subsection{The fundamental-frequency map}\label{subsec:map}

The geodesic equations \eqref{eq:r_geodesic}--\eqref{eq:t_geodesic} take a simple form with the choice of a timelike parameter $\lambda=\int d\tau/\Sigma$ \cite{C1983,M2003}; this decouples \eqref{eq:r_geodesic} and \eqref{eq:theta_geodesic}, and for a bound orbit makes the radial and polar components of motion manifestly periodic with respect to $\lambda$. For the azimuthal and temporal components (whose potentials depend only on $(r,\theta)$), overall rates of evolution may be obtained by averaging \eqref{eq:V_phi} and \eqref{eq:V_t} over many periods of radial and polar motion.

From the radial and polar periods $\Lambda_{r,\theta}$, the average azimuthal rate $\langle d\phi/d\lambda\rangle$ and the average temporal rate $\langle dt/d\lambda\rangle$ (denoted $\Gamma$ by analogy with the Lorentz factor), we may define three angular and dimensionless fundamental frequencies $\Omega_{r,\theta,\phi}$ for the test particle's motion with respect to coordinate time. In terms of $(r_p,r_a,\theta_\mathrm{min})$, these frequencies are written as \cite{DH2004,SF2015}
\begin{equation}
\Omega_r=\frac{2\pi}{\Lambda_r\Gamma},\quad\Omega_\theta=\frac{2\pi}{\Lambda_\theta\Gamma},
\end{equation}
\begin{equation}
\Omega_\phi=\lim_{N\to\infty}\frac{1}{N^2\Lambda_r\Lambda_\theta\Gamma}\int_0^{N\Lambda_r}d\lambda_r\int_0^{N\Lambda_\theta}d\lambda_\theta\,V_\phi,
\end{equation}
where $\Lambda_{r,\theta}$ and $\Gamma$ are given by
\begin{equation}
\Lambda_r=2\int_{r_p}^{r_a}\frac{dr}{\sqrt{V_r}},\quad\Lambda_\theta=4\int_{\theta_\mathrm{min}}^{\pi/2}\frac{d\theta}{\sqrt{V_\theta}},
\end{equation}
\begin{equation}
\Gamma=\lim_{N\to\infty}\frac{1}{N^2\Lambda_r\Lambda_\theta}\int_0^{N\Lambda_r}d\lambda_r\int_0^{N\Lambda_\theta}d\lambda_\theta\,V_t.
\end{equation}
Expressions for $\Omega_{r,\theta,\phi}$ in terms of $(e,\iota,p)$ have been derived by Schmidt \cite{S2002}; these are less compact, but have more utility in practical implementations.

In terms of the fundamental frequencies, the periapsis and Lense--Thirring precession rates are given by $\Omega_\phi-\Omega_r$ and $\Omega_\phi-\Omega_\theta$ respectively. These vanish in the Newtonian limit, where $\Omega_{r,\theta,\phi}$ approach a single orbital frequency $\Omega$ from below, i.e. $\Omega_r\nearrow\Omega_\theta\nearrow\Omega_\phi\nearrow\Omega$. The frequency $\Omega$ is then related to $(e,p)$ by Kepler's third law:
\begin{equation}\label{eq:third_law}
\Omega=\left(\frac{1-e^2}{p}\right)^{3/2}.
\end{equation}
In the AK model, $\Omega$ is associated with the quantity $\dot{\Phi}M$; however, periapsis and Lense--Thirring precession are added on top of $\dot{\Phi}$ via $\dot{\gamma}+\dot{\alpha}$ and $\dot{\alpha}$ respectively, and so $\Omega$ is the lowest frequency by construction. This inconsistency with the relativistic case leads to mismatched frequencies when supplying identical parameters to the two models, since the same value of $p$ in \eqref{eq:third_law} specifies the radial AK frequency while approximating the azimuthal NK frequency. In other words, the frequencies in the AK model are generally too high.

A three-dimensional endomorphism over the AK space of orbits is induced by requiring that the radial, polar and azimuthal frequencies $(\dot{\Phi},\dot{\Phi}+\dot{\gamma},\dot{\Phi}+\dot{\gamma}+\dot{\alpha})$ for any $(e,\iota,p)$ have the same values as the (dimensionful) relativistic frequencies $\omega_{r,\theta,\phi}:=\Omega_{r,\theta,\phi}/M$. We map the parameters $(M,a,p)$ rather than $(e,\iota,p)$ to unphysical values; this gives better results since periapsis and Lense--Thirring precession are more directly determined by the central mass and its rotation respectively. The map $(M,a,p)\mapsto(\tilde{M},\tilde{a},\tilde{p})$ is given implicitly by solving the algebraic system of equations
\begin{equation}\label{eq:Phi_map}
\dot{\Phi}(\tilde{M},\tilde{a},\tilde{p})=\omega_r(M,a,p),
\end{equation}
\begin{equation}\label{eq:gamma_map}
\dot{\gamma}(\tilde{M},\tilde{a},\tilde{p})=\omega_\theta(M,a,p)-\omega_r(M,a,p),
\end{equation}
\begin{equation}\label{eq:alpha_map}
\dot{\alpha}(\tilde{M},\tilde{a},\tilde{p})=\omega_\phi(M,a,p)-\omega_\theta(M,a,p)
\end{equation}
for the unphysical set $(\tilde{M},\tilde{a},\tilde{p})$, which is defined as the root closest to the physical set $(M,a,p)$ with a Euclidean metric on parameter space.

Substituting $(\tilde{M},\tilde{a},\tilde{p})$ for $(M,a,p)$ in the AK model provides an instantaneous correction of its frequencies at any point along the inspiral trajectory $(e(t),\iota(t),p(t))$. In principle, applying the map along the entirety of a fiducial inspiral will keep the AK waveform phase-coherent with relativistic waveforms (generated from that trajectory) until plunge. However, such an inspiral is usually more expensive to compute (as in the case of the NK model), and the additional cost from evaluating the map itself also scales linearly with the number of points sampled along the trajectory. Complications also arise as the compact object approaches the point of plunge, where the fundamental frequencies diverge and the map \eqref{eq:Phi_map}--\eqref{eq:alpha_map} is no longer well-defined.

\subsection{Implementation}\label{subsec:implementation}

In order to retain the main advantage of the AK model, computational costs are kept as low as possible by evaluating the map at a small number of points and relying on independent evolution of the orbital constants over long timescales.  Firstly, the NK phase-space trajectory is generated at and around the specified initial time $t_0$ over a user-defined timescale $T_\mathrm{fit}$, which depends on the radiation-reaction timescale $T_\mathrm{RR}:=M^2/\mu$ and specifies the duration over which the AAK model is calibrated. The timescale $T_\mathrm{fit}$ and number of sample points $N_\mathrm{fit}$ may be adjusted adaptively based on the proximity of the initial point $(e_0,\iota_0,p_0):=(e(t_0),\iota(t_0),p(t_0))$ to plunge; they typically satisfy $0.1T_\mathrm{RR}\lesssim T_\mathrm{fit}\lesssim10T_\mathrm{RR}$ (over which $\iota$ is approximately constant) and $N_\mathrm{fit}\lesssim10$ (to ensure an added computational cost of $\lesssim1\%$). Evaluation of the map at each of the $N_\mathrm{fit}$ points gives a local ``best-fit'' trajectory $(\tilde{M}(t),\tilde{a}(t),e(t),\tilde{p}(t))_\mathrm{fit}$ in the AAK phase space, where the unphysical $(\tilde{M}(t),\tilde{a}(t))_\mathrm{fit}$ change with time.

The global trajectory in the AAK phase space is generated independently from the NK model, and hence more rapidly. From the mapped initial point $(\tilde{M}_0,\tilde{a}_0,e_0,\tilde{p}_0)$ (which also lies on the best-fit trajectory by construction), a global PN trajectory $(\tilde{M},\tilde{a},e(t),\tilde{p}(t))_\mathrm{PN}$ is obtained by evolving $(e(t),\tilde{p}(t))_\mathrm{PN}$ with 3PN $O(e^6)$ expressions given by Sago \& Fujita \cite{SF2015}, while $(\tilde{M},\tilde{a})_\mathrm{PN}$ are left constant. Higher-order 4PN $O(e^6)$ expressions \cite{SF2015} have also been tested, but these seem to result in poorer agreement with NK waveforms (which use fluxes of up to 3PN), possibly due to the known divergence of certain expansions beyond 3PN order \cite{YB2008}.

Sampling the PN trajectory at each of the $N_\mathrm{fit}$ points allows the calculation of $(\tilde{M}(t),\tilde{a}(t),e(t),\tilde{p}(t))_\mathrm{fit}-(\tilde{M},\tilde{a},e(t),\tilde{p}(t))_\mathrm{PN}$ over the duration $T_\mathrm{fit}$. This difference trajectory is fitted to polynomials in time and extrapolated over the lifetime of the inspiral; it is then added to the PN trajectory, giving the final global trajectory $(\tilde{M}(t),\tilde{a}(t),e(t),\tilde{p}(t))$. In the initial AAK implementation, the coefficients of the quadratic fit are given by second-order finite-difference quotients (i.e. $N_\mathrm{fit}=3$), which works well but only for small values of $T_\mathrm{fit}$. The present version uses a quartic least-squares fit, which allows the choice of longer $T_\mathrm{fit}$ and consequently gives better long-term phase agreement with NK waveforms.

\begin{figure}
\centering
\includegraphics[width=\columnwidth]{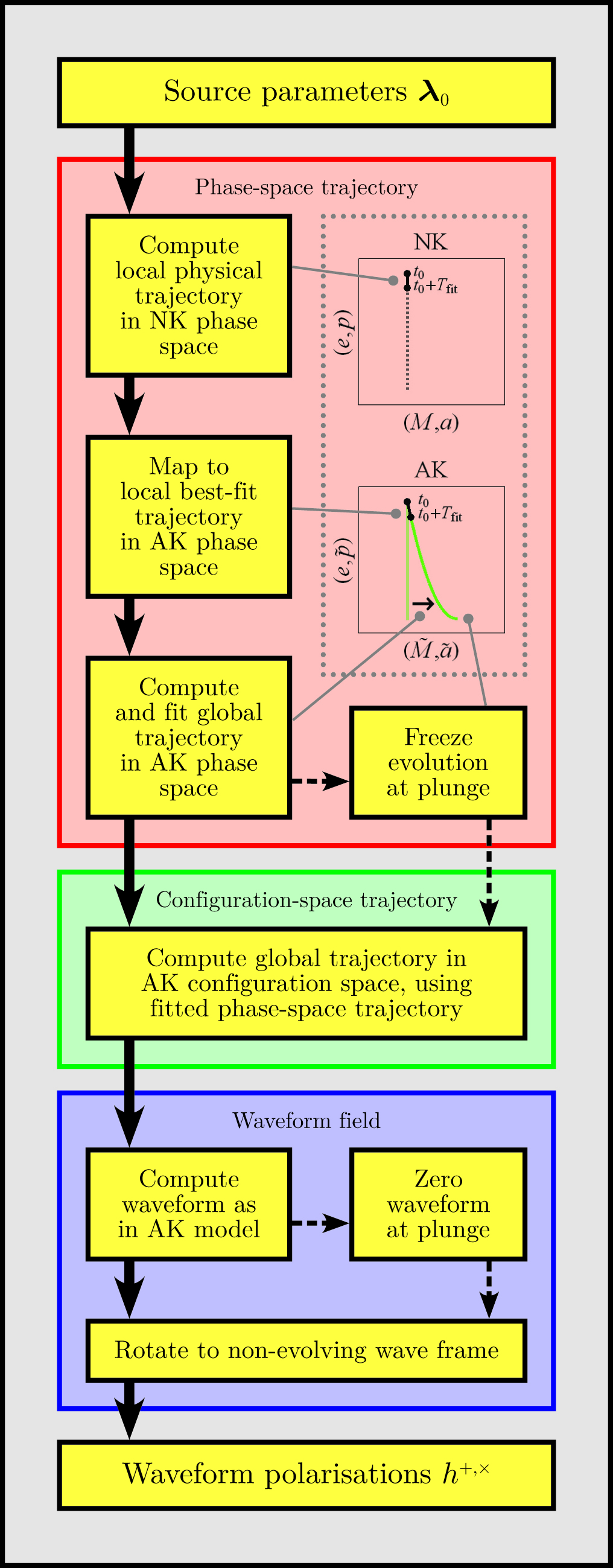}
\caption{Flowchart summary of the AAK waveform model algorithm. Dashed arrows indicate additional steps for an EMRI that plunges within the mission lifetime.}
\label{fig:flowchart}
\end{figure}

After the phase-space trajectory $(\tilde{M}(t),\tilde{a}(t),e(t),\tilde{p}(t))$ has been constructed, the configuration-space evolution of $(\Phi,\gamma,\alpha)$ is performed with the appropriate combinations \eqref{eq:Phi_map}--\eqref{eq:alpha_map} of the fundamental frequencies, given by Sago--Fujita expressions \cite{SF2015} that are consistent at 3PN $O(e^6)$ with the phase-space evolution. The waveform field is then generated as in the AK model. For illustrative purposes, a flowchart summary of the entire AAK algorithm (as presently implemented) is shown in Fig.~\ref{fig:flowchart}.

The augmentations to the AK framework are focused on improving the phase information of its waveforms, since the amplitude of a GW signal is measured far less precisely than its phase. However, the calculation of amplitude in the original AK model ($\mathcal{A}$ in \eqref{eq:PM_a}--\eqref{eq:PM_c}) is a decent approximation since it is based on $\dot{\Phi}M$, which is assigned a value $\approx\Omega_\phi$ that turns out to be correct for this purpose (see discussion around \eqref{eq:third_law}). Hence it is the AAK amplitude that is shifted away from the fiducial NK value through the mapping of frequencies and the unphysical evolution of $\tilde{M}$. A simple adjustment is made to reverse this shift; in \eqref{eq:PM_a}--\eqref{eq:PM_c} for the AAK model, the amplitude is now given by
\begin{equation}
\mathcal{A}=\frac{(\omega_\phi M)^{2/3}\mu}{|\mathbf{R}|},
\end{equation}
where the azimuthal frequency $\omega_\phi$ and the physical black-hole mass $M$ are used in place of $\dot{\Phi}=\omega_r$ and $\tilde{M}$ (i.e. $M$ in the original AK model) respectively.

Finally, a fast method of plunge handling has been added to the present AAK implementation; this feature is useful in general, but especially when generating large numbers of AAK waveform templates for search algorithms. The compact object plunges when its instantaneous orbit along the phase-space trajectory $\mathbf{E}(t)$ becomes unstable, i.e.
\begin{equation}
\frac{\partial^2V_r(r,a,\mathbf{E})}{\partial r^2}\leq\frac{\partial V_r(r,a,\mathbf{E})}{\partial r}=V_r(r,a,\mathbf{E})=0,
\end{equation}
where $V_r$ is given in \eqref{eq:V_r} with $\mathbf{E}=(E,L_z,Q)$. This point is termed the last stable orbit $\mathbf{E}_\mathrm{LSO}$, and is precisely the point at which the discriminant $\mathcal{D}(a,\mathbf{E})$ of the quartic polynomial $V_r(r)$ changes sign from positive (four real roots) to negative (two real roots) \cite{R1922}. Since $\mathcal{D}$ is a simple analytic function of the quartic coefficients, it is computationally trivial to check for stability at every integration step for the phase-space trajectory, provided the evolution is done in terms of $(E,L_z,Q)$.\footnote{The discriminant method is applicable to the NK model, and may speed it up slightly. Currently, the NK implementation precomputes (for the specified value of $a$) an interpolated $p_\mathrm{LSO}$ surface over the relevant region of $(e,\iota)_\mathrm{LSO}$ space, by finding and examining the roots of $V_r(r)$ numerically. It then checks for $p<p_\mathrm{LSO}(e,\iota)$ when generating the phase-space trajectory.}

Plunge detection in the AAK model is far less straightforward than in the other two kludges, partly because the evolution is performed in terms of the quasi-Keplerian orbital parameters, and the computational benefits of the discriminant method are nullified by having to convert $(e,\iota,\tilde{p})\to(E,L_z,Q)$. Furthermore, $(\tilde{M},\tilde{a},\tilde{p})$ are unphysical; the inverse of the map \eqref{eq:Phi_map}--\eqref{eq:alpha_map} is computationally expensive and (more crucially) ill-defined at plunge, and so cannot be used to obtain the physical parameters for stability calculations.

To circumvent these issues, the AAK model uses \eqref{eq:third_law} with $\Omega\approx\Omega_\phi$ to obtain an approximation for the physical parameter $p$. While generating the phase-space trajectory, it checks (between the least- and most-bound orbits \cite{H2001}) the stability of $(e,\iota,p)$ at every radiation-reaction interval $T_\mathrm{RR}$. Once the stability changes across an interval, it then bisects that interval to find $p_\mathrm{LSO}$, and smoothly zeroes the waveform over ten additional orbits with a one-sided Planck-taper window \cite{MRS2010}. The added computational cost associated with this algorithm is $\lesssim1\%$. Although the approximation for $p$ is crude, the phase-space trajectories in the AAK and NK models are generally divergent to begin with, and the plunge points for both models may differ significantly even if a more accurate expression is used.

\subsection{Benchmarking results}\label{subsec:benchmarking}

The specified initial state of an EMRI in the AAK model is described by the intrinsic parameters $(\mu,M,a,e_0,\iota_0,\gamma_0,\psi_0)$ and the extrinsic parameters $(p_0,\theta_S,\phi_S,\theta_K,\phi_K,\alpha_0,D)$, where $D:=|\mathbf{R}|$.\footnote{For a source at cosmological redshift $z$, the values of $D$ and $(\mu,M)$ are replaced with the luminosity distance $D(1+z)$ and the redshifted masses $(\mu(1+z),M(1+z))$ respectively.} In configuration space, transformations from $\mathbf{X}_\mathrm{AAK}=(\psi,\gamma,\alpha)$ to $\mathbf{X}_\mathrm{AK}=(\Phi,\gamma,\alpha)$ and $\mathbf{X}_\mathrm{NK}=(\psi,\chi,\phi)$ are required for a comparison of the three waveform models. We have chosen to specify the quasi-Keplerian true anomaly $\psi$ in the (shared) AAK parameter space, since there is no closed-form expression for $\psi$ in terms of the mean anomaly $\Phi$. The conversion $\psi\to\Phi$ is given by the Keplerian expressions \cite{C2005}
\begin{equation}
\Phi=E-e\sin{E},\quad E=\tan^{-1}\left(\frac{\sqrt{1-e^2}\sin{\psi}}{e+\cos{\psi}}\right),
\end{equation}
where $E$ is known as the eccentric anomaly.

On the other hand, the AAK model retains the AK parameters $(\gamma,\alpha)$; these have explicit meanings in the (intrinsic) $\mathbf{L}$-based coordinate frame
\begin{equation}\label{eq:L-based}
(\hat{\mathbf{x}},\hat{\mathbf{y}},\hat{\mathbf{z}})_\mathbf{L}:=\left(\frac{\hat{\mathbf{L}}\times\hat{\mathbf{S}}}{\mathcal{S}_{\mathbf{L},\mathbf{S}}},\frac{(\hat{\mathbf{S}}\cdot\hat{\mathbf{L}})\hat{\mathbf{L}}-\hat{\mathbf{S}}}{\mathcal{S}_{\mathbf{L},\mathbf{S}}},\hat{\mathbf{L}}\right),
\end{equation}
where the normalisation factor $\mathcal{S}_{\mathbf{L},\mathbf{S}}:=(1-(\hat{\mathbf{L}}\cdot\hat{\mathbf{S}})^2)^{1/2}$ and $\hat{\mathbf{L}}(\alpha)$ is given in ecliptic coordinates by \eqref{eq:L_ecliptic}. The unit position vector of the compact object with respect to \eqref{eq:L-based} is $\hat{\mathbf{r}}_\mathbf{L}=[\cos{(\psi+\gamma)},\sin{(\psi+\gamma)},0]^T$, and a change of basis to the $\mathbf{S}$-based coordinate frame \eqref{eq:S-based} gives
\begin{equation}
\hat{\mathbf{r}}_\mathbf{S}=\mathbf{Q}_\mathbf{S}^T\mathbf{Q}_\mathbf{L}\hat{\mathbf{r}}_\mathbf{L}, 
\end{equation}
where the orthogonal matrices $\mathbf{Q}_\mathbf{S}:=[\hat{\mathbf{x}}|\hat{\mathbf{y}}|\hat{\mathbf{z}}]_\mathbf{S}$ and $\mathbf{Q}_\mathbf{L}:=[\hat{\mathbf{x}}|\hat{\mathbf{y}}|\hat{\mathbf{z}}]_\mathbf{L}$ are formed from the triads in \eqref{eq:S-based} and \eqref{eq:L-based} respectively. It is then straightforward to obtain $(\chi,\phi)$ from $\hat{\mathbf{r}}_\mathbf{S}=[\sin{\theta}\cos{\phi},\sin{\theta}\sin{\phi},\cos{\theta}]^T$, via \eqref{eq:NK_phases}.

As a generic example, we consider a prograde EMRI with redshifted component masses $(\mu,M)=(10^1,10^6)M_\odot$, spin $a=0.5M$, and initial orbital parameters $(e_0,\iota_0,p_0)=(0.1,\pi/6,8.25M)$. The initial semi-latus rectum is chosen such that the compact object plunges approximately one year after entering the LISA band at a representative frequency $f_\mathrm{GW}=2.7\,\mathrm{mHz}$, where $f_\mathrm{GW}$ is defined as twice the azimuthal orbital frequency (i.e. the dominant GW harmonic at low eccentricity). In the AAK model, the fitting timescale and number of sample points are set to $T_\mathrm{fit}\leq10T_\mathrm{RR}$ and $N_\mathrm{fit}\leq10$ respectively, with inequality in the case of adaptive adjustments.

One important result from our comparison studies is that the AK model can lead to an overestimation of SNR if used without modification. This is due to the fact that the frequencies in the AK model are generally too high for any given $(e,\iota,p)$, as mentioned in Sec.~\ref{subsec:AK}. The SNR of a signal $h=h_I+ih_{II}$ is defined as $\rho:=\sqrt{\langle h|h\rangle}$, where $\langle\cdot|\cdot\rangle$ is the standard matched-filtering inner product (on the space of finite-length time series) between two waveforms, i.e. \cite{CF1994}
\begin{equation}
\langle a|b\rangle=2\int_0^\infty df\,\frac{\tilde{a}^*(f)\tilde{b}(f)+\tilde{a}(f)\tilde{b}^*(f)}{S_n(f)}.
\end{equation}
Throughout this paper, the noise power spectral density $S_n(f)$ is taken to be a LISA noise model for the L6A5 (six links, five-million-kilometre arms) configuration known as classic LISA, assuming the original mission requirements and including confusion noise from the foreground of Galactic white-dwarf binaries \cite{KEA2016}.

\begin{figure}
\centering
\includegraphics[width=\columnwidth]{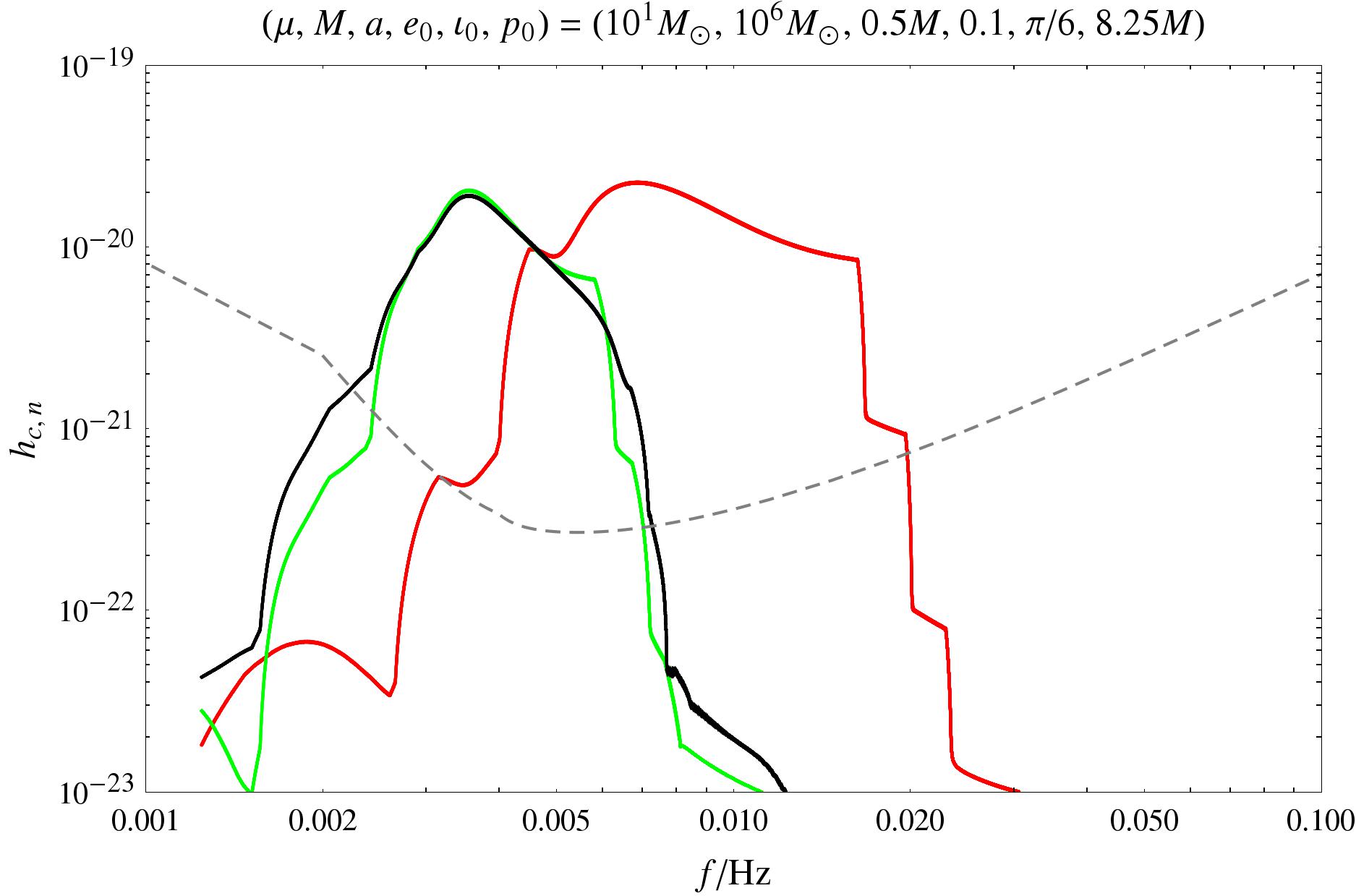}
\caption{Characteristic strain for year-long AK (red), AAK (green) and NK (black) signals from the example EMRI, along with the noise amplitude (dashed) for the LISA configuration L6A5. A moving-average filter has been applied to the $h_c$ curves, such that oscillations are smoothed out for ease of visualisation while the overall spectral profile is preserved.}
\label{fig:characteristic_strain}
\end{figure}

At a luminosity distance of 5\,Gpc, an NK signal (sampled at 0.2\,Hz) from the example EMRI described above has a one-year SNR of $\rho=30.8$. When using the AAK model to generate the signal, a comparable value of $\rho=32.1$ is obtained. However, an AK signal from the same EMRI has $\rho=57.8$. To illustrate this, we consider the characteristic strain $h_c$ of a signal and the noise amplitude $h_n$; these are given respectively by \cite{MCB2015}
\begin{equation}
h_c(f)=2f|\tilde{h}(f)|^2,\quad h_n(f)=\sqrt{fS_n(f)},
\end{equation}
such that
\begin{equation}
\rho^2=\int_{-\infty}^\infty d(\ln{f})\,\left(\frac{h_c(f)}{h_n(f)}\right)^2.
\end{equation}
With these definitions, the area between $h_c$ and $h_f$ on a log--log plot gives an indication (but not an approximation) of SNR, and allows the relative detectability of signals to be estimated. The characteristic strain for the three signals and the LISA noise amplitude are shown in Fig.~\ref{fig:characteristic_strain}, where the excess power in the AK signal at higher frequencies is evident, along with the consequent boost to SNR. This error is likely to persist for $M\gtrsim10^6M_\odot$, but may be mitigated for less massive central black holes as the maxima of the three $h_c$ curves are blueshifted past the minimum of $h_n$.

For the purposes of this paper (where the NK model is taken as fiducial), the phase accuracy of the AK and AAK models is assessed by how well their waveforms overlap with NK waveforms. The overlap $\mathcal{O}$ between two waveforms $a$ and $b$ is defined as
\begin{equation}\label{eq:overlap}
\mathcal{O}(a|b):=\frac{\langle a|b\rangle}{\sqrt{\langle a|a\rangle\langle b|b\rangle}},
\end{equation}
which takes the value of one for identical waveforms and zero for orthogonal waveforms. In \cite{CG2015}, the overlaps $\mathcal{O}(h_\mathrm{AK}|h_\mathrm{NK})$ and $\mathcal{O}(h_\mathrm{AAK}|h_\mathrm{NK})$ over two and six months are computed for the example EMRI, as well as for the same source with (i) $\mu=10^0M_\odot$, (ii) $a=0.8M$ and (iii) $e_0=0.5$. The AK and initial AAK models have virtually identical computation times $\tau$, and are both quicker than the NK model with typical speed-up factors of $\sigma:=1-\tau/\tau_\mathrm{NK}\approx0.9$ (except in the case of $e_0=0.5$, where $\sigma\approx0.4$). However, the AAK model yields overlaps that are consistently higher, and by 2--3 orders of magnitude in most cases.

\begin{figure}
\centering
\includegraphics[width=\columnwidth]{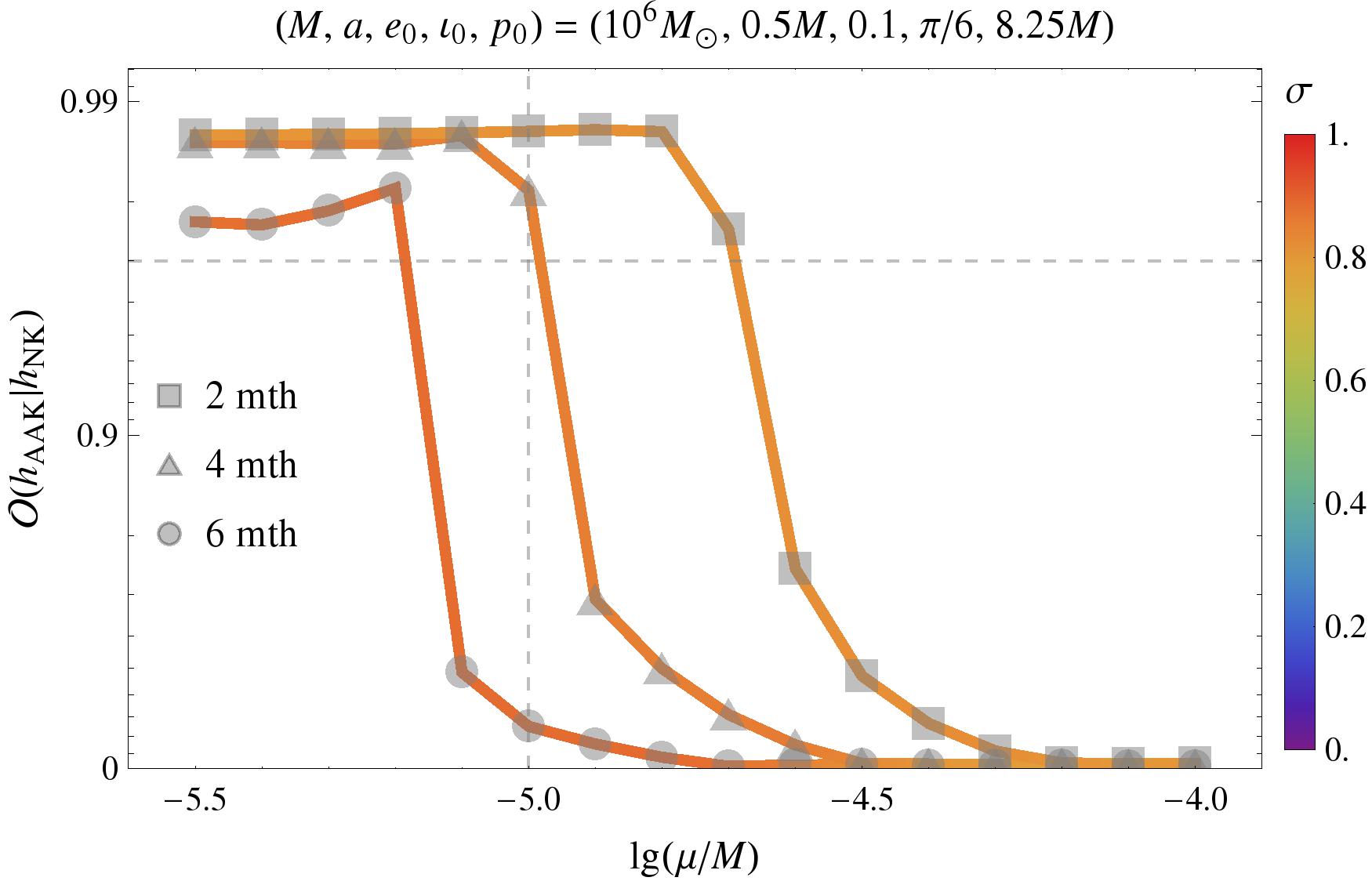}
\caption{Two- to six-month overlaps between AAK and NK waveforms for generic EMRIs with varying compact-object mass. The computational speed-up $\sigma$ of the AAK over the NK is coded by colour. Vertical dashed lines correspond to the example EMRI, while horizontal ones indicate the standard minimal-match value of 0.97 for template banks.}
\label{fig:benchmark_mass}
\end{figure}

\begin{figure}
\centering
\includegraphics[width=\columnwidth]{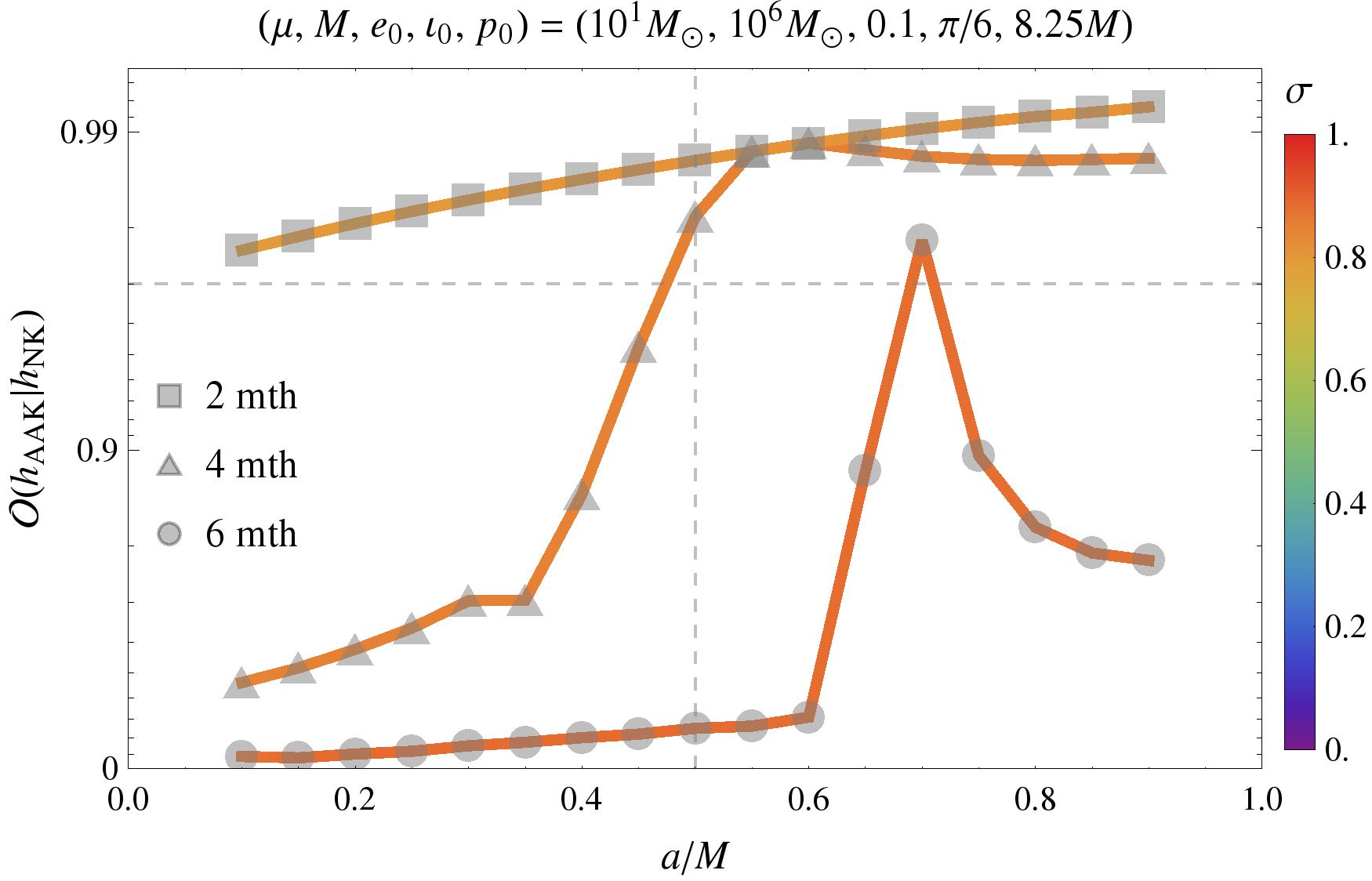}
\caption{Overlaps and computational speed-up as in Fig.~\ref{fig:benchmark_mass}, but for generic EMRIs with varying black-hole spin.}
\label{fig:benchmark_spin}
\end{figure}

\begin{figure}
\centering
\includegraphics[width=\columnwidth]{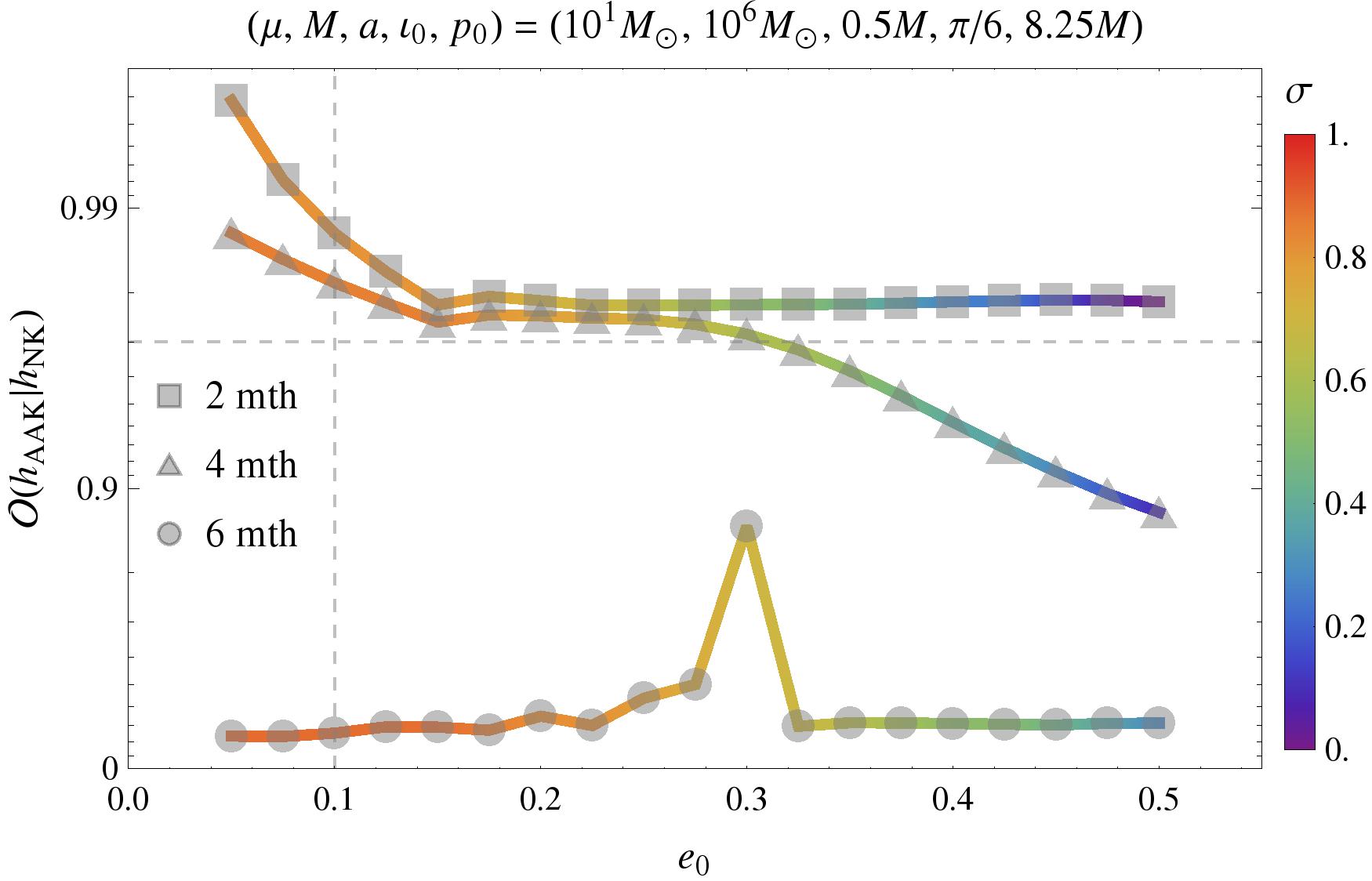}
\caption{Overlaps and computational speed-up as in Fig.~\ref{fig:benchmark_mass}, but for generic EMRIs with varying initial eccentricity.}
\label{fig:benchmark_eccentricity}
\end{figure}

The speed-up factors for the AAK model are preserved by the present implementation, while its overlaps are increased across the board due to the enhanced fitting algorithm. Most notably, there is substantial improvement for EMRIs with higher initial eccentricities, although this is partly attributable to the use of a higher sample rate than that in \cite{CG2015}. The overlap and timing performance of the AAK model across 2--6 months with varying compact-object mass, black-hole spin and initial eccentricity is summarised by the plots in Figs~\ref{fig:benchmark_mass}--\ref{fig:benchmark_eccentricity}.

In Fig.~\ref{fig:benchmark_mass}, the overlap $\mathcal{O}(h_\mathrm{AAK}|h_\mathrm{NK})$ is computed for the example EMRI, as well as for the same source with $0.5\leq\lg{(\mu/M_\odot)}\leq2$. The minimum value of $\mu\approx3M_\odot$ is chosen since the upper bound of $10T_\mathrm{RR}$ for $T_\mathrm{fit}$ exceeds six months for a less massive compact object, and so the overlaps in that regime will not show significant improvement (even the six-month overlap is already $>0.97$). We also do not consider intermediate-mass black holes with $\mu>100M_\odot$. Instead of choosing $p_0$ such that the EMRI plunges after one year (as done in \cite{CG2015}), we consider fixed $p_0=8.25M$ in this analysis; this leads the overlaps to degrade at larger rather than smaller mass ratios $\mu/M$. The two-month overlaps are $>0.97$ up to $\mu\approx20M_\odot$, for which plunge occurs at around 5.6 months. There is greater speed-up over the NK model for longer waveform durations as expected, but all values of $\sigma$ are $\gtrsim0.8$.

Overlaps for the example EMRI with varying spin $0.1\leq a/M\leq0.9$ are shown in Fig.~\ref{fig:benchmark_spin}. Again, all values of $\sigma$ are $\gtrsim0.8$, with greater speed-up for longer waveform durations. For fixed $p_0$, prograde EMRIs with lower spin start closer to plunge, and so the overlap values generally increase along with $a$. There appears to be an opposing effect at higher spin (possibly due to the additional degrees of freedom for error from fitting $\tilde{M}$ and $\tilde{a}$ in the AAK model) that causes a fall-off in the four- and six-month overlaps for $a\gtrsim0.6$. The two-month overlaps are $>0.97$ across the full range of considered spins, which is perhaps unsurprising since the upper bound of $10T_\mathrm{RR}$ for $T_\mathrm{fit}$ is also two months for a $(10^1,10^6)M_\odot$ EMRI. However, we note here that while the phase accuracy of the AAK model may be arbitrarily increased in principle by taking $T_\mathrm{fit}>10T_\mathrm{RR}$, the computational requirement that $N_\mathrm{fit}\lesssim10$ will likely reduce the quality of the trajectory fit at early times.

The effect of varying eccentricity for the example EMRI is illustrated in Fig.~\ref{fig:benchmark_eccentricity}. We consider initial eccentricities $0.05\leq e_0\leq0.5$, since there turns out to be no speed-up over the NK model ($\sigma\approx0$) when generating a two-month AAK waveform with $e_0=0.5$. This is an important limitation of the Peters--Mathews approximation for waveform generation (see discussion in footnote \ref{foot:modes}), and will have to be addressed if the AAK model is to be useful in searches for high-eccentricity EMRIs. Nevertheless, the two- and even four-month overlaps are $>0.97$ for $e_0\lesssim0.3$, with $\sigma\gtrsim0.5$. As in the case of Fig.~\ref{fig:benchmark_spin}, there is a peak in the six-month overlap; this is probably due to variance in the fit for $e$, and is unlikely to carry any fundamental significance.

Finally, although parameter estimation with AAK waveforms is not the focus of this paper, we summarise here the results of a Fisher matrix calculation for a year-long AAK signal from the example EMRI considered throughout Sec.~\ref{subsec:benchmarking}. The Fisher information matrix $\boldsymbol{\Gamma}$ for a GW signal $h$ parametrised by $\boldsymbol{\lambda}$ is given by \cite{CF1994}
\begin{equation}
\boldsymbol{\Gamma}_{ij}=\left\langle\frac{\partial h}{\partial\boldsymbol{\lambda}_i}\big|\frac{\partial h}{\partial\boldsymbol{\lambda}_j}\right\rangle,
\end{equation}
where $\boldsymbol{\lambda}=(\mu,M,\mathbf{S},\mathbf{E},\mathbf{X},\mathbf{R})$ for EMRIs. The parameter-estimation errors $\Delta\boldsymbol{\lambda}$ due to Gaussian noise have the normal distribution $\mathcal{N}(\mathbf{0},\boldsymbol{\Gamma}^{-1})$ in the case of high SNR, and so the root-mean-square errors in the general case can be approximated as
\begin{equation}
\Delta\boldsymbol{\lambda}_i\approx\sqrt{(\boldsymbol{\Gamma}^{-1})_{ii}}.
\end{equation}

For our AAK signal normalised to an SNR of $\rho=30$, we find that the masses and spin can be measured to within the fractional errors
\begin{equation}
\Delta(\ln{\mu})\approx4\times10^{-5}
\end{equation}
\begin{equation}
\Delta(\ln{M})\approx2\times10^{-5},
\end{equation}
\begin{equation}
\Delta(\ln{(a/M)})\approx4\times10^{-5}.
\end{equation}
These errors are roughly an order of magnitude better than the corresponding values for the AK model \cite{BC2004}, and are more comparable to those cited for the NK model \cite{HG2009} (although the latter consider a circular, equatorial EMRI with $a=0.9M$). They are also consistent with (the lower end of) the values reported in the L2/L3 mission proposal for ESA's Cosmic Vision programme \cite{AEA2013}, where the AK model was used but with a modified plunge criterion.

\section{Data analysis application: Semi-coherent detection searches}\label{sec:application}

GW detection is usually achieved via a template bank search, in which a large set of signal templates $h_i$ is compared against the noisy detector data $s$. A single coherent integral of template and data is calculated for each template in the bank and used as a detection statistic, i.e. a detection is claimed if any of these values exceeds a predetermined threshold. For EMRI detection, such a procedure is hampered by the extremely large number of templates needed to cover the parameter space; a previous estimate \cite{GEA2004} put this number at $N_\mathrm{bank}\sim10^{40}$. In that same work, a computationally viable alternative was suggested: a semi-coherent search that involves splitting the time series data into $N$ segments, searching each segment separately with a smaller template bank, then combining the results to obtain a new detection statistic. Inevitably, such a search is less sensitive than the (computationally prohibitive) fully coherent search.

The minimum feasible number of segments for a semi-coherent EMRI search was estimated in \cite{GEA2004} by considering the computational resources anticipated to be available when LISA flies; it was found that $N\gtrsim100$ for a mission lasting $\sim10^8\,\mathrm{s}$, giving segments of length $\Delta T\lesssim10^6\,\mathrm{s}$. In this work, we assume that computational resources are not a limiting factor and instead consider the loss in performance of the semi-coherent search compared to the fully coherent search, estimated as a function of $N$. For a waveform model to safely be used in a semi-coherent search, it must remain phase-accurate over the duration of each segment, and so the different dephasing times in the AK and AAK models will determine the maximum $\Delta T$ for each model and the corresponding loss in performance. This section will discuss whether either model is sufficiently accurate for real LISA data analysis.

For an EMRI template bank $\{h_i\,|\,i=1,2,\ldots,N_\mathrm{bank}\}$, we may define the fully coherent detection statistic for each template as
\begin{equation}
\rho_i:=\frac{\langle s|h_i\rangle}{\sqrt{\langle h_i|h_i\rangle}},
\end{equation}
i.e. the template SNR. If the measured data consists solely of Gaussian noise ($s=n$), it is straightforward to show that $\rho_i\sim\mathcal{N}(0,1)$ (each template SNR is distributed as a zero-mean normal random variable). In the presence of an EMRI signal ($s=h_i+n$), we have $\rho_i\sim\mathcal{N}(A,1)$ for the corresponding template, where $A:=\langle h_i|h_i\rangle^{1/2}$ is the signal amplitude. A detection is claimed if any $\rho_i$ exceeds a predetermined threshold $\rho_*$, which may be set by fixing a desired false-alarm probability $P_F$:
\begin{eqnarray}\label{eq:FAR}
&&P_F(\rho_*)=\int_{\rho_*}^\infty d\rho_i\,\frac{1}{\sqrt{2\pi}}\exp{\left(-\frac{\rho_i^2}{2}\right)}\nonumber\\
&\implies&\rho_*(P_F)=\sqrt{2}\,\mathrm{erfc}^{-1}(2P_F).
\end{eqnarray}

In practice, $\rho_i$ is not used as a detection statistic because it is computationally cheaper to analytically search over several extrinsic parameters (rather than having to generate templates that vary in those parameters). Each template may be written as
\begin{equation}
h_i(t)=Au_i(t-t_{c})\exp(i\phi_{c}),
\end{equation}
where the normalised templates $u_i$ satisfy $\langle u_i|u_i\rangle=1$. The waveform amplitude $A$, time-of-arrival $t_c$ and phase offset $\phi_c$ are extrinsic parameters that may be searched over for each $u_i$ at negligible additional cost \cite{BCV2003}.

We may now define a fully coherent phase-maximised detection statistic as
\begin{equation}
\tilde{\rho}_i:=\max_{\phi_c}\langle s|u_i\rangle,
\end{equation}
i.e. $\rho_i$ maximised over $\phi_{c}$. Unlike the template SNR, $\tilde{\rho}_i$ is not normally distributed; in the no-signal case $s=n$, it follows a Rayleigh distribution with unit scale parameter, and has the probability density function \cite{BCV2003}
\begin{equation}
f_0(\tilde{\rho}_i)=\tilde{\rho}_i\exp{\left(-\frac{\tilde{\rho}_i^2}{2}\right)},\quad\tilde{\rho}_i\geq0.
\end{equation}
If $s=h_i+n$, then $\tilde{\rho}_i$ for the corresponding template follows a Rice distribution with unit scale parameter and offset parameter $A$; its probability density function is \cite{BCV2003}
\begin{equation}
f_1(\tilde{\rho}_i,A)=\tilde{\rho}_i\exp{\left(-\frac{\tilde{\rho}_i^2+A^2}{2}\right)}I_0(A\tilde{\rho}_i),\quad\tilde{\rho}_i\geq0,
\end{equation}
where $I_0$ denotes the order-zero modified Bessel function of the first kind.

A detection threshold $\tilde{\rho}_*$ for the new statistic may be chosen in similar fashion to \eqref{eq:FAR}:
\begin{eqnarray}
&&P_F(\tilde{\rho}_*)=\int_{\tilde{\rho}_*}^\infty d\tilde{\rho}_i\,f_0(\tilde{\rho}_i)\nonumber\\
&\implies&\tilde{\rho}_*(P_F)=-2\ln{P_F}.
\end{eqnarray}
In the following analysis, we consider a desired false-alarm probability of $10^{-3}$ for the entire template bank. By approximating $\{\tilde{\rho}_i\,|\,i=1,2,\ldots,N_\mathrm{bank}\}$ as a set of independent random variables, we may simply reduce this value by a trials factor of $N_\mathrm{bank}$ to obtain $P_F$ (the desired false-alarm probability for a single template). We may also assume that the time-of-arrival has been maximised over (e.g. using fast Fourier transforms \cite{BCCS1998}), which incurs an additional trials factor of $N_\mathrm{time}\sim10^8$ (for $t_c$ offsets of 1\,s). Hence we set
\begin{equation}\label{eq:fixed_FAR}
P_F=\frac{10^{-3}}{N_\mathrm{bank}N_{\mathrm{time}}}=10^{-51}.
\end{equation}

\begin{figure}
\centering
\includegraphics[width=\columnwidth]{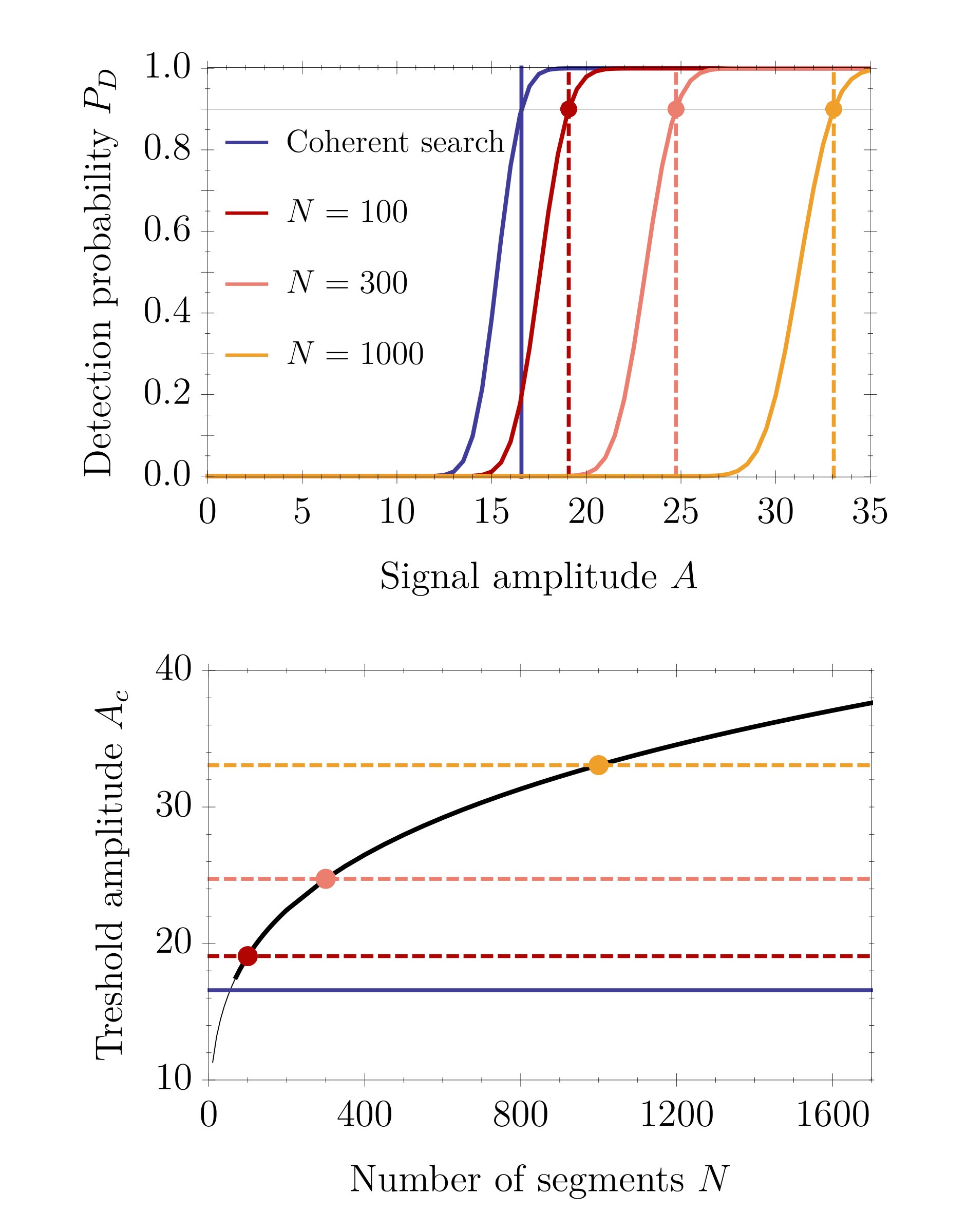}
\caption{Top panel: Detection probability $P_D$ as a function of signal amplitude $A$ for fully coherent search and semi-coherent search with three different numbers of segments $N$. Vertical lines indicate the threshold amplitude at which $P_D=0.9$. Bottom panel: Threshold amplitude $A_{c}$ as a function of $N$ for semi-coherent search; horizontal lines correspond to the vertical lines in the top panel. For $N\lesssim70$, the distribution of the semi-coherent detection statistic $\Upsilon_i$ can no longer be approximated as Gaussian and the black curve becomes unreliable (indicated on the plot by the use of a thinner stroke).}
\label{fig:semi_coherent}
\end{figure}

With $\tilde{\rho}_*$ fixed by \eqref{eq:fixed_FAR}, we now consider the detection probability
\begin{equation}\label{eq:P_D}
P_D(A)=\int_{\tilde{\rho}_*}^\infty d\tilde{\rho}_i\,f_1(\tilde{\rho}_i,A),
\end{equation}
i.e. the probability in the presence of a signal that $\tilde{\rho}_i>\tilde{\rho}_*$ for the corresponding template. The detection probability as a function of $A$ is shown as the blue curve in the top panel of Fig.~\ref{fig:semi_coherent}, and the critical signal amplitude needed to achieve $P_D=0.9$ is $A_c\approx17$. This threshold is the SNR required for an EMRI to be detected by an idealised fully coherent search; for a semi-coherent search, $A_c$ will increase with $N$ as detection sensitivity is lost.

In a semi-coherent search where the data is split into $N$ segments of equal length, the phase-maximised detection statistic $\tilde{\rho}_i$ is calculated as before but for each segment $j$, and the semi-coherent detection statistic is then taken to be the quadrature sum
\begin{equation}
\Upsilon_i:=\sum_{j=1}^N(\tilde{\rho}_{i,j})^2.
\end{equation}
Although each of the $\tilde{\rho}_{i,j}$ follows a Rayleigh/Rice distribution, the distribution of $\Upsilon_i$ is analytically intractable. However, in the limit $N\to\infty$, the central limit theorem guarantees that $\Upsilon_i$ will be normally distributed. We find empirically that for $N\gtrsim70$, the distribution of $\Upsilon_i$ is well approximated as Gaussian; the mean and variance in the respective absence/presence of a signal are given by
\begin{equation}
(\mu_0,\sigma^2_0)=(N\mu_k,2N\sigma_k^2),
\end{equation}
\begin{equation}
(\mu_1,\sigma^2_1)=(N\mu_k+A^2,2N\sigma_k^2+4A^2),
\end{equation}
with $\mu_k\approx2.00$ and $\sigma_k\approx1.45$. The functional forms of $(\mu_1,\sigma^2_1)$ are motivated by the corresponding expressions for a noncentral chi-squared distribution.

Using this normal approximation for the distribution of $\Upsilon_i$, the detection threshold $\Upsilon_*$ is set by the fixed false-alarm probability \eqref{eq:fixed_FAR} as
\begin{eqnarray}
&&P_F(\Upsilon_*)=\int_{\Upsilon_*}^\infty d\Upsilon_i\,\frac{1}{\sqrt{2\pi\sigma^2_0}}\exp{\left(-\frac{(\Upsilon_i-\mu_0)^2}{2\sigma^2_0}\right)}\nonumber\\
&\implies&\Upsilon_*(P_F)=\mu_0-\sqrt{2\sigma^2_0}\,\mathrm{erf}^{-1}(2P_F-1),
\end{eqnarray}
while the detection probability is given by
\begin{eqnarray}\label{eq:P_Dsemi}
P_D(A)&=&\int_{\Upsilon_*}^\infty d\Upsilon_i\,\frac{1}{\sqrt{2\pi\sigma^2_1}}\exp{\left(-\frac{(\Upsilon_i-\mu_1)^2}{2\sigma^2_1}\right)}\nonumber\\
&=&\frac{1}{2}\left(1+\mathrm{erf}\left(\frac{\mu_1(A)-\Upsilon_*}{\sqrt{2\sigma^2_1(A)}}\right)\right).
\end{eqnarray}
Along with the fully coherent expression \eqref{eq:P_D} in the top panel of Fig.~\ref{fig:semi_coherent}, the semi-coherent detection probability \eqref{eq:P_Dsemi} is also plotted for several values of $N$; the threshold signal amplitude $A_c$ for which $P_D=0.9$ in each case is indicated by the corresponding vertical line.

In the bottom panel of Fig.~\ref{fig:semi_coherent}, $A_c$ is shown as a function of $N$; through a reduced chi-squared fit, we find that this relationship is well approximated as the power law 
\begin{equation}\label{eq:power_law}
A_c\approx6.57\times N^{0.235}
\end{equation}
in the range $1\leq N\lesssim10^4$ (well beyond the range plotted in Fig.~\ref{fig:semi_coherent}). As the number of segments is increased, the threshold SNR at which an EMRI can be detected increases, i.e. the search becomes less sensitive and a greater number of events will be missed. However, \eqref{eq:power_law} taken at face value suggests that for $N\lesssim55$, the semi-coherent search is more sensitive than the fully coherent search ($A_c<17$). This conclusion is obviously incorrect, and is due to the breakdown of the normal approximation for the distribution of $\Upsilon_i$ at $N\lesssim70$ (indicated on the plot by the transition to a thinner black curve).

Since the AK model typically dephases over a few hours ($\sim10^4\,\mathrm{s}$), employing AK waveforms in a semi-coherent detection search would necessitate the use of $N\sim10^4$ segments to cover the full mission lifetime of $\sim10^8\,\mathrm{s}$. This incurs an intolerable loss of performance; the threshold SNR at which an EMRI can be detected would be raised from 17 to 59, where this number is estimated by using the power law \eqref{eq:power_law} and dividing by an assumed model accuracy of 0.97 over the dephasing time. On the other hand, only $N\sim10^2$ segments are required for the AAK model with its typical dephasing time of two months ($\sim10^6\,\mathrm{s}$), which would raise the threshold SNR from 17 to 20. It was also found previously in \cite{GEA2004} that $N\sim10^2$ is the minimum number of segments needed for a computationally feasible semi-coherent search.

\begin{table*}
\begin{center}
\begin{tabular}{|c|c|C|C|C|C|C|C|C|C|C|C|}
\hline
&&\multicolumn{10}{c|}{Number of events in mass range}\\\cline{3-12}
Plunge&Population&\multicolumn{2}{c}{$M_{10}<5$}&\multicolumn{2}{|c}{$5<M_{10}<5.5$}&\multicolumn{2}{|c}{$5.5<M_{10}<6$}&\multicolumn{2}{|c}{$M_{10}>6$}&\multicolumn{2}{|c|}{Total}\\\cline{3-12}
criterion&model&AAK&AK&AAK&AK&AAK&AK&AAK&AK&AAK&AK\\\hline
\multirow{12}{*}{Schwarzschild}&M1&20&0&240&10&110&10&10&0&380&20\\
&M2&30&0&190&10&70&10&0&0&290&10\\
&M3&20&0&310&10&510&40&40&10&880&50\\
&M4&70&0&280&20&80&20&0&0&440&40\\
&M5&0&0&10&0&20&0&0&0&30&0\\
&M6&20&0&270&10&210&10&20&0&520&30\\
&M7&230&0&2190&60&1040&100&60&10&3530&180\\
&M8&0&0&30&0&10&0&0&0&50&0\\
&M9&20&0&210&0&110&10&10&0&350&20\\
&M10&30&0&240&10&100&10&10&0&370&10\\
&M11&0&0&0&0&1&0&0&0&1&0\\
&M12&230&10&2420&70&1730&130&180&30&4560&230\\\hline
\multirow{12}{*}{Kerr}&M1&20&0&260&10&230&10&80&10&590&30\\
&M2&20&0&210&0&160&10&50&10&440&20\\
&M3&10&0&360&10&1000&60&240&50&1620&120\\
&M4&50&0&300&20&140&30&30&10&520&70\\
&M5&0&0&10&0&40&0&40&10&90&10\\
&M6&20&0&300&10&430&30&200&50&960&80\\
&M7&190&0&2390&60&2110&150&730&120&5420&330\\
&M8&0&0&30&0&30&0&10&0&70&0\\
&M9&20&0&230&0&160&10&30&0&430&20\\
&M10&30&0&240&10&100&10&10&0&370&10\\
&M11&0&0&0&0&1&0&0&0&1&0\\
&M12&190&0&2700&60&3710&210&1830&410&8440&690\\\hline
\end{tabular}
\end{center}
\caption{Number of EMRI events when using AAK and AK waveforms in a semi-coherent detection search, for the 12 astrophysical EMRI-population models considered in \cite{BEA2017} (M1--12) and two different plunge criteria (Schwarzschild and Kerr innermost stable circular orbits). Events are further divided into four black-hole mass bins, where $M_{10}:=\lg{(M/M_\odot)}$. Event counts are rounded to the nearest 10 for all models except M11, where they are rounded to the nearest 1.}
\label{tab:Nevent}
\end{table*}

In moving from the AK model to the AAK model, the lowering of the threshold SNR from 59 to 20 has a profound effect on the likely number of detected EMRIs. Tab.~\ref{tab:Nevent} shows how many events LISA would expect to observe, using the thresholds appropriate for both kludges. Numbers are reported for the 12 astrophysical EMRI-population models that were recently considered in \cite{BEA2017}. These populations make different assumptions about the characteristics of the MBHs that play hosts to EMRIs, the typical masses of the compact objects involved, and the amount of mass that the central black holes can accrete; we refer the reader to \cite{BEA2017} for full details. Event SNRs are computed with two different plunge criteria: either the innermost stable circular orbit for a Schwarzschild black hole (as in the original AK paper \cite{BC2004}), or that for a Kerr black hole. As argued in \cite{BEA2017}, these two assumptions should give values that bracket the true SNR. Detected events are further divided into four mass ranges for the central black hole.

We see that some detections would still be expected if the AK model is used to analyse LISA data, although significantly fewer (by at least an order of magnitude) than if the more faithful AAK model is employed. Moreover, we would lose the events at the edges of the distribution (i.e. in the lowest and highest mass ranges), and so the astrophysical information provided by the LISA EMRI population would be reduced. The AAK model, on the other hand, would find a significant number of events under most of the population models, with its required threshold SNR of 20 for a semi-coherent search being compatible with the thresholds that have traditionally been assumed when assessing LISA's capability for EMRI detection. Results for a fully coherent search are not shown in Tab.~\ref{tab:Nevent}, but reducing the threshold SNR from 20 to 17 would increase the event rate by only a modest amount (50\% or less); furthermore, it is unlikely that the computing power needed for such a search will be available even in the mid-2030s. 

\section{Conclusion}

We have developed an augmented variant of Barack \& Cutler's widely used EMRI kludge waveform model \cite{BC2004}; the new AAK model retains the speed of its predecessor, while matching the phase evolution of more accurate but slower kludges over a significant fraction of the inspiral. With the latest implementation of the model released online at \href{https://github.com/alvincjk/EMRI_Kludge_Suite}{github.com/alvincjk/EMRI_Kludge_Suite} as part of a kludge software suite, AAK waveforms will hopefully see widespread use in the next round of mock LISA data challenges.

One existing deficiency in the AAK model is the ill-defined nature of the fundamental frequency map \eqref{eq:Phi_map}--\eqref{eq:alpha_map} at the last stable orbit, due to the divergent Kerr frequencies; this complicates both plunge detection and the specification of orbital parameters at plunge. Another limitation is that the mode-sum approximation \eqref{eq:mode_sum} becomes more expensive than the quadrupole formula itself at high eccentricities ($e_0\gtrsim0.5$). Work is ongoing to resolve these issues, and to make the model as streamlined and robust as possible for the mock data challenges.

Regardless, the present implementation of the AAK model shows significantly improved accuracy over larger fractions of the inspiral as compared to the initial AAK implementation \cite{CG2015} (which itself extends the dephasing time of the original AK model from hours to months). The two-month overlaps of AAK waveforms with NK waveforms for a variety of EMRIs with different compact-object mass, black-hole spin and initial eccentricity are typically increased from $\lesssim0.95$ (in the initial implementation) to $>0.97$ by the techniques presented in this paper. Computational efficiency for the present implementation is also retained to within 1\% of the AK and initial AAK models, with all analytic kludges able to generate generic waveforms 5--15 times more quickly than the NK model (except when $e_0\gtrsim0.3$).

We have also considered the performance of the AAK model in a data analysis application: the semi-coherent EMRI detection search proposed in \cite{GEA2004}. An analytic estimate is provided for the threshold SNR needed to detect an EMRI with a false-alarm probability of $\sim10^{-3}$ when using a bank of $\sim10^{40}$ templates. This threshold is 17 for a fully coherent search with templates that are 97\% accurate over the inspiral lifetime, although such a search is likely to be out of reach computationally. If the AK model is used in a semi-coherent search, the threshold rises to 59 and around 95\% fewer EMRI events will be detected (under various population models); however, a semi-coherent AAK search requires a lower threshold of 20 due to the model's longer dephasing times, and will yield at least an order of magnitude more events than the AK search while remaining computationally feasible. This suggests that unlike the AK model, the AAK model might realistically be employed in actual LISA data analysis without much loss in detection sensitivity.

\begin{acknowledgments}
We thank Stanislav Babak, Leor Barack, Christopher Berry and Anthony Lasenby for helpful discussions and/or comments on the manuscript. AJKC's work was supported by the Cambridge Commonwealth, European and International Trust. CJM has received funding from the European Union's Horizon 2020 research and innovation programme under the Marie Sk\l odowska-Curie grant agreement No.~690904, and from the STFC Consolidator Grant No.~ST/L000636/1.
\end{acknowledgments}

\bibliographystyle{unsrt}
\bibliography{references}
\end{document}